\documentclass[twocolumn, tighten]{aastex701}
\received{, 2025}
\revised{, 2025}
\accepted{202X}
\submitjournal{ApJ}
\graphicspath{{./}{Figures/}}
\shorttitle{Inner Galaxy gradients}
\shortauthors{Sales-Silva et al.}

\begin{document}

\title{Chemical radial gradients for the bulge-bar stellar populations from the APOGEE survey}

\author[0000-0003-0636-7463]{J. V. Sales-Silva}
\affil{Observat\'orio Nacional/MCTI, R. Gen. Jos\'e Cristino, 77,  20921-400, Rio de Janeiro, Brazil}
\email[show]{joaovsaless@gmail.com}

\author[0000-0001-6476-0576]{K. Cunha}
\affiliation{Steward Observatory, University of Arizona Tucson AZ 85719}
\affil{Observat\'orio Nacional/MCTI, R. Gen. Jos\'e Cristino, 77,  20921-400, Rio de Janeiro, Brazil}
\email{kcunha@arizona.edu}

\author[0000-0002-0134-2024]{V. V. Smith}
\affiliation{NOIRLab, Tucson AZ 85719}
\email{verne.smith@noirlab.edu}

\author{S. Daflon}
\affiliation{Observat\'orio Nacional/MCTI, R. Gen. Jos\'e Cristino, 77,  20921-400, Rio de Janeiro, Brazil}
\email{daflon@on.br}

\author{D. Souto}
\affiliation{Departamento de F\'isica, Universidade Federal de Sergipe, Av. Marechal Rondon, S/N, 49000-000 S\~ao Crist\'ov\~ao, SE, Brazil}
\email{diogosouto@academico.ufs.br}

\author[0000-0002-0151-5212]{R. Guerço}
\affil{Universidad Cat\'olica del Norte, N\'ucleo UCN en Arqueolog\'ia Gal\'actica - Inst. de Astronom\'ia, Av. Angamos 0610, Antofagasta, Chile}
\affiliation{Observat\'orio Nacional/MCTI, R. Gen. Jos\'e Cristino, 77,  20921-400, Rio de Janeiro, Brazil}
\email{guercorafael@gmail.com }

\author{V. Loaiza-Tacuri}
\affiliation{Departamento de F\'isica, Universidade Federal de Sergipe, Av. Marechal Rondon, S/N, 49000-000 S\~ao Crist\'ov\~ao, SE, Brazil}
\email{vloatac@gmail.com}

\author{A. Queiroz}
\affil{Instituto de Astrofísica de Canarias (IAC), E-38205 La Laguna, Tenerife, Spain}
\affil{Universidad de La Laguna (ULL), Departamento de Astrofísica, 38206 La Laguna, Tenerife, Spain}
\email{anna.barbara@iac.es}

\author{C. Chiappini}
\affil{Leibniz-Institut f{\"u}r Astrophysik Potsdam (AIP), An der Sternwarte 16, 14482 Potsdam, Germany}
\email{cristina.chiappini@aip.de}

\author{I. Minchev}
\affil{Leibniz-Institut f{\"u}r Astrophysik Potsdam (AIP), An der Sternwarte 16, 14482 Potsdam, Germany}
\email{iminchev1@gmail.com}

\author{S. R. Majewski}
\affil{Department of Astronomy, University of Virginia, Charlottesville, VA 22904-4325, USA}
\email{srm4n@virginia.edu}

\author{B. Barbuy}
\affil{Universidade de São Paulo, IAG, Rua do Matão 1226, Cidade Universitária, São Paulo 05508-090, Brazil}
\email{b.barbuy@iag.usp.br}

\author{D. Bizyaev}
\affil{Apache Point Observatory and New Mexico State University, P.O. Box 59, Sunspot, NM 88349-0059, USA}
\affil{Sternberg Astronomical Institute, Moscow State University, Moscow, Russia}
\email{dmbiz@apo.nmsu.edu}

\author{Jos\'e G. Fern\'andez-Trincado}
\affil{Universidad Cat\'olica del Norte, N\'ucleo UCN en Arqueolog\'ia Gal\'actica - Inst. de Astronom\'ia, Av. Angamos 0610, Antofagasta, Chile}
\affil{Universidad Cat\'olica del Norte, Departamento de Ingenier\'ia de Sistemas y Computaci\'on, Av. Angamos 0610, Antofagasta, Chile}
\email{jose.fernandez@ucn.cl}

\author[0000-0002-0740-8346]{Peter M.~Frinchaboy}
\affiliation{Department of Physics and Astronomy, Texas Christian University, TCU Box 298840 Fort Worth, TX 76129, USA }
\email{p.frinchaboy@tcu.edu}

\author{S. Hasselquist}
\affil{Space Telescope Science Institute, Baltimore, MD, USA}
\email{shasselquist@stsci.edu}

\author{D. Horta}
\affil{Institute for Astronomy, University of Edinburgh, Royal Observatory, Blackford Hill, Edinburgh EH9 3HJ, UK}
\email{dhortadarrington@gmail.com}

\author[0000-0002-4912-8609]{Henrik J\"onsson}
\affil{Materials Science and Applied Mathematics, Malm\"o University, SE-205 06 Malm\"o, Sweden}
\email{henrik.jonsson@mau.se}

\author{T. Masseron}
\affil{Instituto de Astrofísica de Canarias (IAC), E-38205 La Laguna, Tenerife, Spain}
\affil{Universidad de La Laguna (ULL), Departamento de Astrofísica, 38206 La Laguna, Tenerife, Spain}
\email{thomas.masseron@iac.es}

\author{N. Prantzos}
\affil{Institute d'Astrophysique de Paris, UMR 7095 CNRS \& Sorbonne Universit\'e, 98 bis Blvd Arago, Paris  75014  France}
\email{prantzos@iap.fr}

\author{R. P. Schiavon}
\affil{Astrophysics Research Institute, Liverpool John Moores University, Liverpool, L3 5RF, UK}
\email{R.P.Schiavon@ljmu.ac.uk}

\author{M. Schultheis}
\affil{Université Côte d’Azur, Observatoire de la Côte d’Azur, CNRS, Laboratoire Lagrange, Bd de l’Observatoire, CS 34229, 06304 Nice Cedex 4, France}
\email{mathias.schultheis@oca.eu}

\author{M. Zoccali}
\affil{Instituto de Astrofísica, Pontificia Universidad Católica de Chile, Vicunã Mackenna 4860, Macul, Casilla 306, Santiago 22, Chile}
\email{mzoccali@uc.cl}

\begin{abstract}

The Milky Way bulge-bar is composed of multiple populations. Using chemical and kinematical planes, we segregate six populations in a bulge-bar sample observed by the APOGEE survey: two with bar-driven orbits, two with eccentric orbits, and two with low-eccentricity orbits, each composed of low- and high-[Mg/Fe] stars. Our sample spans $-2.0\lesssim$[Fe/H]$\lesssim+0.5$ and Galactocentric distance $R_{Gal}$ $<6$ kpc. We use chemical abundances from APOGEE DR17 for the elements Mg, Si, Ca, Al, K, Mn, Co, Ni, and Fe, and from the BAWLAS catalog for Ce and Nd. We find that the low- and high-[Mg/Fe] stars with low-eccentricity orbits, which exhibit chemical and orbital characteristics similar to those of the low- and high-[$\alpha$/Fe] disks, display slightly negative and positive metallicity gradients, respectively. This result for the low-[Mg/Fe] low-eccentricity stars indicates a break in the global thin disk metallicity gradient. The high eccentricity populations with both low- and high-[Mg/Fe] show approximately flat metallicity gradients. In general, the [X/H] gradients of all elements for all populations follow Fe, except for the neutron-capture elements Ce and Nd. For all elements, the high-[Mg/Fe] bar population shows a much steeper positive [X/H] gradient than the nearly flat gradient for the low-[Mg/Fe] bar stars. The positive [X/H] gradients observed among our high-[Mg/Fe] bar stars probably reflect an age variation along the peanut structure. This interpretation agrees with the N-body simulations. Such steep positive gradients have also been reported in some high-redshift (z$\sim$4--10) galaxies. 
\end{abstract}

\keywords{Galaxy:evolution -- Galaxy:abundances -- Galaxy:bulge -- Field stars:abundances}


\section{Introduction}

The distribution of metals in the Milky Way (MW) reflects the history of the Galaxy's formation and evolution, as shaped by the processes of our Galaxy's interactions (e.g., accretion), stellar migration, star formation, stellar nucleosynthetic yields, and gas flows. This panorama places chemical abundance gradients as valuable observable constraints on models of the Milky Way's chemical evolution. 

Radial abundance trends of the thin disk are well established in the literature through the analysis of open clusters \citep[e.g.,][]{Magrini2023,Palla2024,Carbajo-Hijarrubia2024,Otto2025}, Cepheids \citep[e.g.,][]{Genovali2014,daSilva2023}, and field stars (e.g., O-B stars, \citealt{daflon2004}, \citealt{Braganca2019}; red giants, \citealt{Anders2017}, \citealt{Johnson2025}; planetary nebula, \citealt{Stanghellini2018}), with the radial metallicity gradient of the thin disk varying roughly between $\sim$$-$0.02  
to $\sim$$-$0.08 dex kpc$^{-1}$, depending on the sample and population analyzed. Different types of objects or populations probe distinct gradient epochs from Galaxy formation. Models from \citet{Minchev2018} find that the interstellar medium (ISM) radial metallicity gradient is steeper at the time of disk formation and flattens as the Galaxy evolves.
The negative value for the thin disk metallicity gradient corroborates an inside-out disk formation scenario (where the innermost regions are assumed to have formed faster than the outermost ones) adopted in chemical evolution models \citep[e.g.,][]{Chiappini2001,Kubryk2015,Schonrich2017,Frankel2019,Prantzos2023}. On the other hand, the thick disk stellar population (morphologically defined\footnote[1]{The thick disk can be defined in different ways: morphologically through decomposition of vertical density profiles or at a fixed height above the disk midplane, kinematically, chemically (high-$\alpha$ sequence), or as the old part of the disk. For a detailed discussion of the different definitions of the thick disk, see \citet{Martig2016}.}) shows a flat radial metallicity trend \citep[e.g.,][]{Cheng2012, Eilers2022}, or inverted and positive \citep[e.g.,][]{Carrell2012,Anders2014,Hayden2014,Sun2024}, as seen also in chemodynamical simulations \citep[e.g.,][]{Li2017,Minchev2014,Miranda2016}. This gradient may represent a turbulent thick disk formation characterized by intense star formation and is consistent with the radial migration of old thick disk stars \citep{Cheng2012}. 
In particular, using APOGEE DR16 RGB stars, \citet{Eilers2022} investigated the radial chemical gradient in the low- and high-$\alpha$ populations from the disk for 20 chemical species, including the inner region of the Galaxy. They found no chemical gradients for high-$\alpha$ stars, no abundance variations due to bar, and that the radial metallicity gradient from the inner galaxy is negative and varies continuously in the low-$\alpha$ population. Additionally, \citet{Ratcliffe2023b} analyzed the gradient in low- and high-$\alpha$ disk populations in bins of ([Fe/H], [Mg/Fe]), finding near-zero slopes for many [X/Fe] ratios, such as for Al, alpha, and iron-peak elements. 
In the extragalactic context, metal radial gradients have been observed in most disk galaxies, indicating that abundances decrease outward and may also flatten over time \citep[e.g.,][]{Li2025a}. Positive gradients have also been observed in some galaxies \citep[e.g.,][]{Troncoso2014,Xin2022}. 

A bar with a boxy/peanut morphology and an X-shaped structure \citep[e.g.,][]{McWilliam2010,Zoccali2017,Wegg2013,Ness2016} characterizes the bulge of the Milky Way, composed of stars spanning a broad metallicity range ($-$1.5 $<$ [Fe/H] $<$ 0.5, \citealt{Barbuy2018}) and mostly old \citep[with possibly only a small fraction of the more metal-rich stars being younger than 5 Gyr,][]{Bensby2017}. Like the local disk, the MW bulge also features stars with high and low [$\alpha$/Fe] ratios \citep[e.g.,][]{Rojas-Arriagada2019,Queiroz2020,Queiroz2021}. Observations of its stellar populations and Galactic evolutionary models posit the formation of the MW bulge as a combination of violent, early-universe processes such as accretion of small Galactic fragments (imprinted in the classical-spheroidal bulge component) and dynamical secular evolution of the disk (traced by the pseudo-bar-boxy-peanut component).

In general, the radial abundance gradients for stellar populations in the inner Galaxy ($R_{Gal}<5$ kpc) are still little explored and literature results are generally limited to the analysis of Fe and $\alpha$ elements \citep[e.g.,][]{Minniti1995,Smartt2001,Rich2007,Zoccali2008,Gonzalez2013,Queiroz2021}, indicating that in regions closer to the Galactic center the stars are more metal-poor or metal-rich depending on whether the sample is near or far from the Galactic plane \citep{Queiroz2020,Queiroz2021}. 
Probing chemical gradients in the inner Galaxy is challenging in part due to high extinction caused by the dust present in this region, but also due to crowded fields.
In addition, fields are composed of a complex mixture of stellar populations: stars from the bulge-bar, in-situ disk stars or disk stars born outside the bulge region but have migrated inward, the halo passing through the inner region, globular and open clusters (or lost stars from these clusters, such as N-rich stars), as well as stars accreted by our Galaxy and located in the central region \citep[e.g., stars from Heracles,][]{Horta2021}.
The high extinction present in the inner Galaxy limits optical observations to a few windows, such as Baade's window \citep[e.g.,][]{McWilliam1994,Cunha2006,Hill2011}, or well above the plane \citep[e.g.,][]{Sestito2023}. In recent years, with the advent of Gaia and the availability of high-resolution near-infrared spectra, detailed multidimensional (chemical, positional, and kinematic) analyzes of stars in the inner Galaxy have become possible \citep[e.g.,][]{GarciaPerez2018,Zasowski2019,Fernandez-Trincado2019,Rojas-Arriagada2019,Queiroz2021,Khoperskov2025,Nepal2025}. Using Gaia astrometric data and the high-resolution spectroscopic survey APOGEE \citep[The Apache Point Observatory Galactic Evolution Experiment;][]{Majewski2017} in the $H$ band (1.514–1.696 $\mu$m), \citet{Khoperskov2025} found that there is no universal metallicity gradient for the bulge of the Milky Way. To understand the formation of the inner Milky Way and its multiple populations more fully, it is essential to analyze the specific chemical gradients of each population separated via a chemo-dynamical analysis. 

Using APOGEE data, \citet{Queiroz2021} performed a detailed orbital-chemical analysis of the stellar populations in the Galactic bulge-bar, segregating the different populations: stars from the bar, from the spheroidal pressure-supported bulge component, and from the inner thin and inner thick disk. This allowed, for example, the discovery that the stellar population with a high probability of being in the boxy-peanut bulge is composed of stars with high and low [$\alpha$/Fe] ratios \citep{Queiroz2021}.

Based on the sample of \citet{Queiroz2021} of the Galactic bulge-bar, in this study, we estimate the radial gradients in the different populations in the inner Galaxy for the $\alpha$-elements Mg, Si, Ca, the odd-Z elements Al, K, the Fe-peak elements Mn, Co, Ni, and Fe, and the heavy elements Ce, and Nd. 
Studying the radial gradients for distinct elements will reveal their sensitivity to the different timescales of element production and to variation in star formation histories. In Section 2 of this paper, we detail the sample and the methodologies adopted to determine the abundances. The chemo-kinematical segregation of the bulge-bar populations is discussed in Section 3. Subsequently (Section 4), we show the chemical gradient for the main populations present in our sample and discuss our findings, comparing them with the gradient from distinct populations and analyzing Galactic evolution scenarios that fit our results. Concluding remarks on our results are found in Section 5.

\section{Sample and chemical abundances} 

We estimated the radial gradients for the bulge-bar stellar populations using the chemical abundances from Data Release 17 \citep[DR17,][]{Abdurrouf2022} of the APOGEE survey \citep{Majewski2017}, which provides the analysis of high-resolution spectra (R$\sim$22,500) in the NIR \citep[1.514 to 1.696 microns,][]{Wilson2019} for 650,000 stars. APOGEE observations were carried out in both hemispheres, using the 2.5 and 1.0 m telescopes at Apache Point Observatory \citep[New Mexico, USA,][]{Gunn2006,Holtzman2010} and the 2.5 m Irenee Du Pont telescope at Las Campanas Observatory \citep[La Serena, Chile,][]{Bowen1973}. \citet{Nidever2015} and \citet{Abdurrouf2022} described the APOGEE data reduction pipeline. The APOGEE Stellar Parameters and Chemical Abundance Pipeline \citep[ASPCAP,][]{GarciaPerez2016} and the APOGEE line list \citep{Smith2021} were used to determine the chemical abundances and atmospheric parameters in DR17. In this study, we used the chemical abundances of Mg, Si, Ca, Al, K, Mn, Co, Ni, and Fe from DR17 to estimate the gradients.

We also analyzed the neutron-capture elements Ce and Nd. In the APOGEE spectra, these elements have weak absorption lines of Ce II and Nd II that overlap with other spectral lines \citep{Cunha2017,Hasselquist2016}, hindering the performance of the ASPCAP pipeline in estimating their abundances. We used the Ce and Nd abundances from the BACCHUS Analysis of Weak Lines in the APOGEE (BAWLAS) catalog \citep{Hayes2022}, which carefully reanalyzed the APOGEE spectra to estimate precise abundances for these elements using the BACCHUS code \citep[][]{Masseron2016}. However, the BAWLAS sample is restricted to APOGEE red giants having calibrated ASPCAP stellar parameters between 3500 K $< $T$_{eff} <$ 5000 K and log g $< $3.5 and  high signal-to-noise spectra (S/N $>$ 150). 

We used the bulge-bar sample from \citet{Queiroz2021}, as done by \citet{Sales-Silva2024} in the Ce and Nd analysis for the bulge-bar stars. \citet{Queiroz2021} selected only stars with high S/N ($>$50) and an excellent spectrum fit using the ASPCAP pipeline (ASPCAP\_CHI2$<$25).
\citet{Queiroz2021} utilized a Bayesian tool, StarHorse \citep{Santiago2016,Queiroz2018}, to derive distances, extinctions, and other astrophysical parameters of stars based on APOGEE, Gaia, and photometric (2MASS, PanSTARRS-1, and AllWISE) data. This analysis allowed access to the 6D phase space of the APOGEE stars with excellent precision, with an uncertainty of around 7\% in distance. With this mapping, \citet{Queiroz2021} selected 26,500 stars in the central region of the Galaxy with Galactocentric Cartesian coordinates contained within $|$X$_{Gal}|<$ 5 kpc, $|$Y$_{Gal}|<$3.5 kpc, and $|$Z$_{Gal}|<$1.0 kpc. To clean the bulge-bar sample of the most apparent disk and halo stars, Queiroz et al. used the reduced proper-motion diagram, a tool to segregate distinct kinematical populations \citep[e.g.,][]{Chiu1980,Faherty2009,Smith2009}, and found $\sim$8000 stars that belong to the Galactic bulge-bar. In this study, using this cleaned bulge-bar sample, we analyze the radial chemical abundance gradients of the following elements: Mg, Si, Ca, Al, K, Mn, Co, Ni, and Fe. We then cross-correlate this sample with the BAWLAS catalog, resulting in a sample of 2098 stars with which we investigate radial gradients of the elements Ce and Nd. 

The mean errors and their respective standard deviations of the atmospheric parameters for our sample are $\sigma_{logg}$= 0.03$\pm$0.01, $\sigma_{T_{eff}}$= 7$\pm$2 K and $\sigma_{[Fe/H]}$= 0.011$\pm$0.002. The average uncertainty of the abundances of the other elements is $\sigma_{Mg}$ =0.015$\pm$0.004, $\sigma_{Si}$= 0.018$\pm$0.005, $\sigma_{Ca} $= 0.018$\pm$ 0.006, $\sigma_{Al} $= 0.031$\pm$0.008, $\sigma_{K}$= 0.046$\pm$0.011, $\sigma_{Mn} $= 0.020$\pm$ 0.007, $\sigma_{Co}$ = 0.051$\pm$0.026, $\sigma_{Ni}$ = 0.017$\pm$0.006, $\sigma_{Ce}$= 0.064$\pm$0.034 and $\sigma_{Nd}$ = 0.087$\pm$0.051.

We investigate the potential systematic effects of atmospheric parameters on the chemical abundances of DR17. Our sample comprises giant stars (0.0 $<$ log g $<$ 3.0), with the bulk of the log g values between 0.5 and 2.5. The selection of red giants displays only minor trends in both Teff and log g as functions of Rgal, partially due to our rather restricted ranges in galactocentric distance (R$_{Gal}\approx$0-5 kpc) and longitude towards the Galactic center. We also select only elements with high-precision measurements for giants in DR17 that follow the trends expected from the literature. Our sample constraints, with restricted atmospheric parameters and precise abundance estimates, reduce the effect of atmospheric parameter systematics on both the DR17 abundances and the resulting abundance gradient estimates. Furthermore, potential concerns about empirical abundance corrections \citep[e.g.,][]{Eilers2022,Sit2024} are mitigated by the absence of a strong correlation between log g, distance, and abundances in our sample.


\section{Chemo-kinematical segregation of the bulge-bar populations}

\begin{figure*}
\centering
	\includegraphics[width=1.0\textwidth]{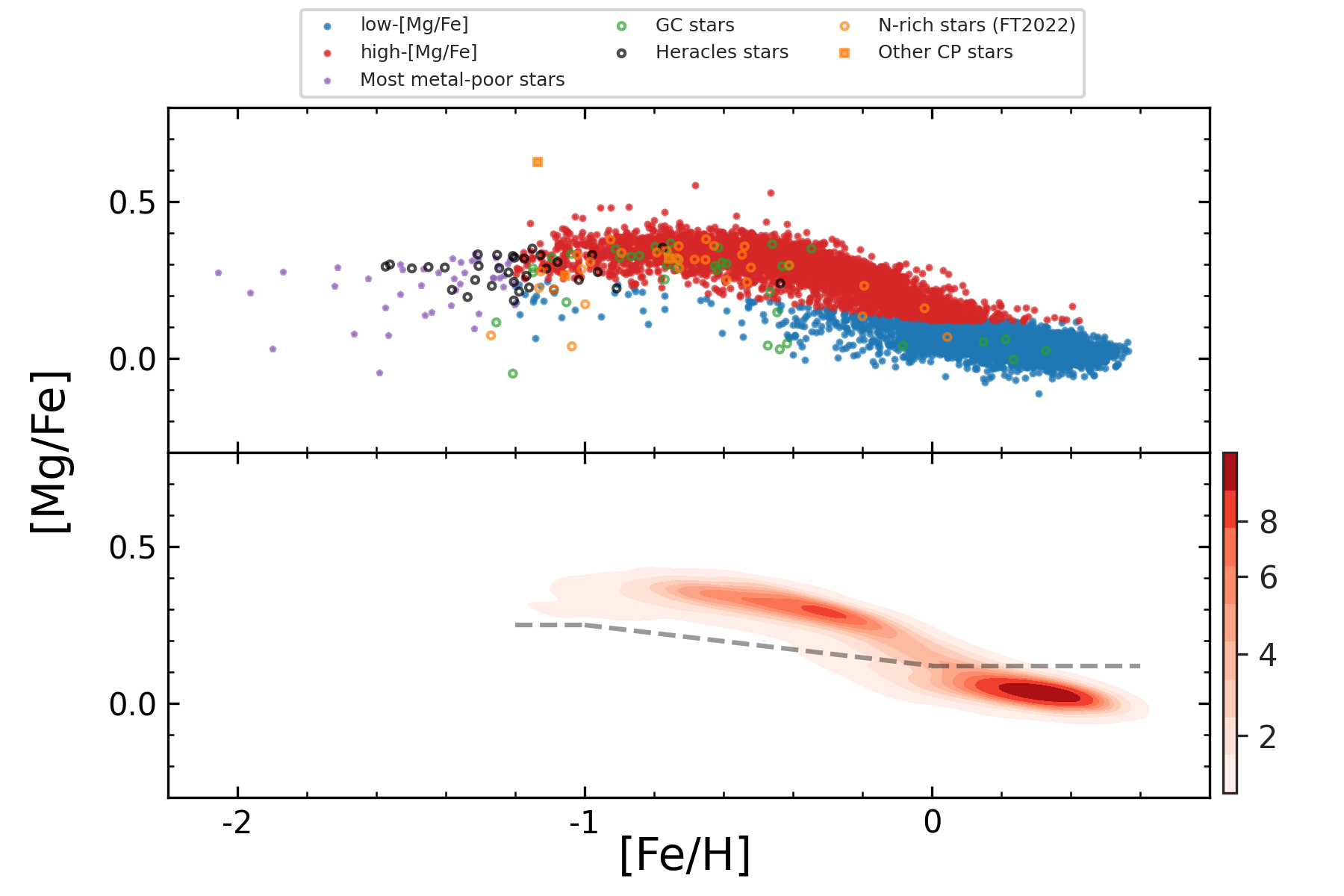}
    \caption{The [Mg/Fe]-[Fe/H] plane for the studied inner Galaxy sample. The top panel shows the low [Mg/Fe] sequence in blue and the high [Mg/Fe] sequence in red. Stars from our sample that have been identified as being from globular clusters, members of the Heracles substructure, N-rich, or chemically peculiar are marked with different symbols. The most metal-poor stars are also marked.  
    The bottom panel shows the relative density distribution in this chemical plane, color-coded according to the probability density function. The dashed line separates the low- and high-[Mg/Fe] populations, as described in the text.}
    \label{MgFe_FeH}
\end{figure*}

The inner galaxy is known to be composed of multiple populations, as indicated by the multi-peaked metallicity distribution function in the bulge \citep[e.g.,][]{Ness2013,Zoccali2017,Rojas-Arriagada2017,Johnson2022}. 
Recently, \citet{Nepal2025} revealed some characteristics of the different populations present in the inner region of the Galaxy using data from \citet{Queiroz2021} and from the Gaia DR3 Radial Velocity Spectrograph. For example, \citet{Nepal2025} found that the metallicity distribution of the bulge spheroidal population peaks at around $-$0.7 dex, and is dominated by a high-[$\alpha$/Fe] population.
We separated the populations in the inner Milky Way by using chemical planes and kinematical decomposition within the $|$Z$_{max}|$-$ecc$ plane, but we also investigated the presence of known stellar populations in our sample using literature catalogs of chemically peculiar stars (N-rich stars), globular cluster stars \citep{VasilievBaumgardt2021,Schiavon2024}, and possibly ex-situ stars (\citealt{Horta2021,Nepal2025}), as done in \citet{Sales-Silva2024}.

First, we analyzed the presence of globular cluster stars within our bulge-bar sample using catalogs of \citet{VasilievBaumgardt2021} and \citet{Schiavon2024}.
\citet{VasilievBaumgardt2021} cataloged the probable member stars of 170 GCs using data from Gaia Early Data Release 3 (EDR3), while \citet{Schiavon2024} cataloged the stars of all known Galactic GCs observed by APOGEE in DR17 based on position, proper motion, and radial velocity. We cross-matched the probable members (P $>$ 0.7) of the GCs from both of these catalogs against our reduce-proper-motion-diagram-segregated bulge-bar sample and found 37 stars that are possible members of globular clusters. We also investigated the presence of N-rich stars, a chemically peculiar population known to be present in the inner galaxy. \citet[FT2022]{Fernandez-Trincado2022} provided a catalog containing 412 N-rich stars in the APOGEE survey, 34 of which are in our sample. We also separated the three chemically peculiar stars detected in \citet{Sales-Silva2024} that are not in the FT2022 sample. Finally, we found 33 stars within our bulge-bar sample that belonged to the Heracles substructure, from cross-matching our bulge-bar sample with the Heracles stars in \citet{Horta2021}.  
In addition, as in our previous study (\citealt{Sales-Silva2024}), we consider stars with [Fe/H]$< -$1.2 as a separate bulge-bar population, naming these stars as the most metal-poor bulge-bar stars, which are likely related to the inner stellar halo or proto-galactic population \citep{Rix2022,Ardern-Arentsen2024,Nepal2025}.

\begin{figure*}
\centering
	\includegraphics[width=1.0\textwidth]{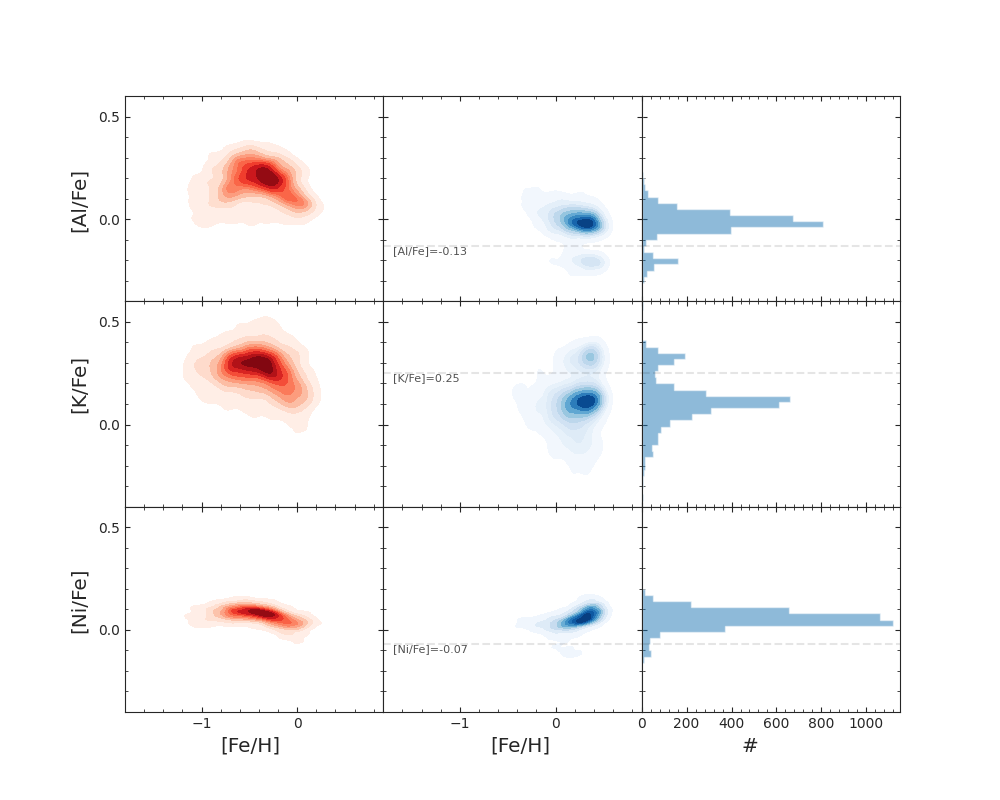}
    \caption{Distribution for the elements Al, K, and Ni, color-coded according to the probability density function. The left and center panels show the red and blue distributions for the high- and low-[Mg/Fe] populations, respectively. The dashed lines in the center panel separate the multiple chemical populations of the low-[Mg/Fe] bulge-bar stars. The right panel presents the histogram for the low-[Mg/Fe] stars.}
    \label{multiple_chemical_populations}
\end{figure*}

\begin{figure*}
\centering
	\includegraphics[width=1.0\textwidth]{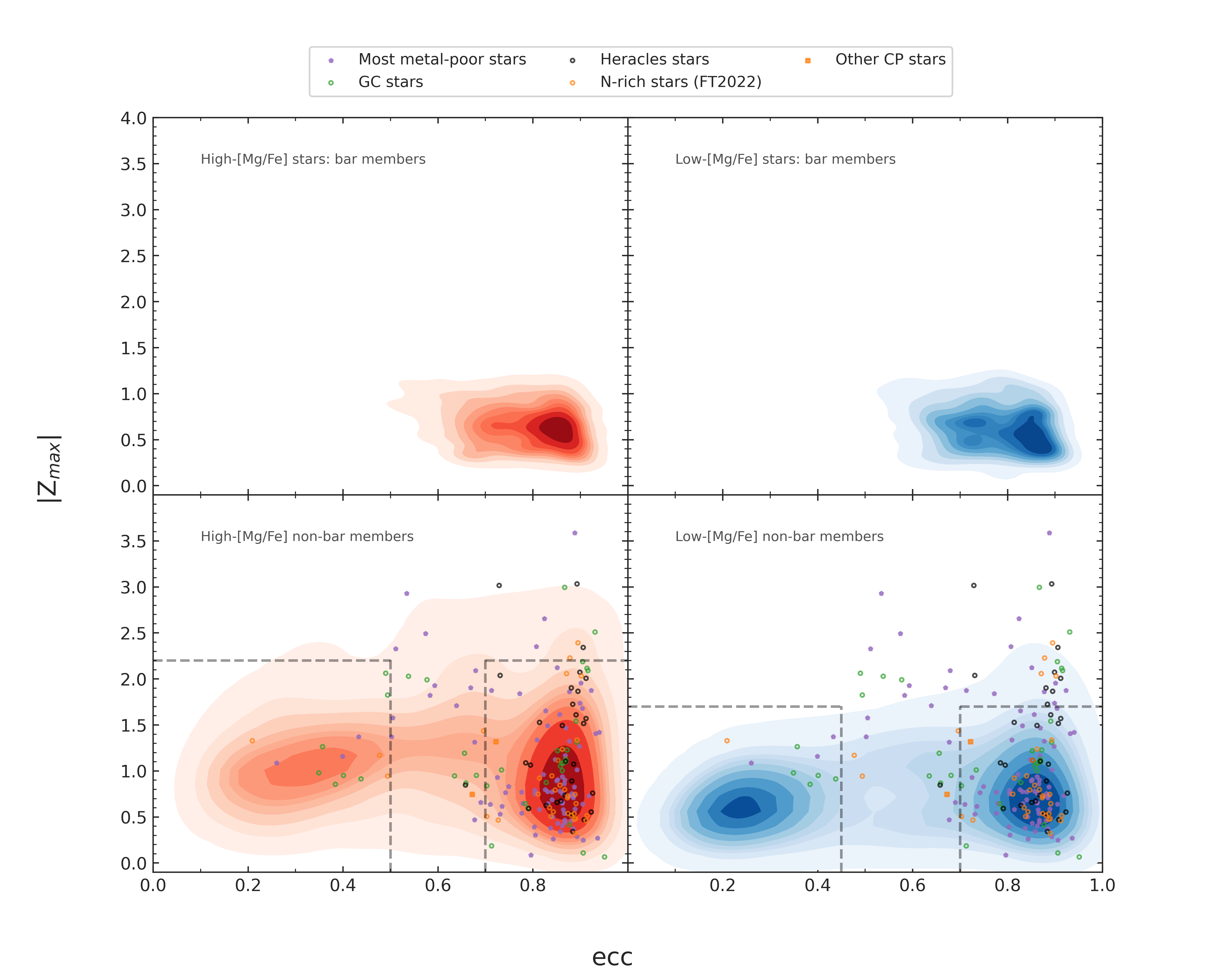}
    \caption{$|Z_{max}|$-eccentricity ($ecc$) plane for the bulge-bar sample. Upper panels relating to the low-(right) and high-[Mg/Fe] (left) bar-member stars, and in the lower panels for the low-(right) and high-[Mg/Fe] (left) stars in the inner Galaxy non-bar member stars. The dashed lines limit the high and low eccentricity populations for the low-(right) and high-[Mg/Fe] (left) stars. Other stars not included in the bulge-bar sample (GC, Heracles, chemically peculiar and most metal-poor stars), are shown as filled and open symbols.}
    \label{kinematic_populations}
\end{figure*}

In Figure \ref{MgFe_FeH} we segregate the low- and high-[Mg/Fe] bulge-bar populations, using the same criteria as in \citet{Sales-Silva2024}. Looking at the density distribution (bottom panel of Figure \ref{MgFe_FeH}), two sequences stand out in the [Mg/Fe]-[Fe/H] plane: one population centered close to solar [Mg/Fe] values ($\sim$0.05 dex) and high [Fe/H] ($\sim$0.3 dex), and the other with high [Mg/Fe] ratios and low [Fe/H]. From now on, we designate these two sequences as the low-[Mg/Fe] and high-[Mg/Fe] populations. The dashed line in the bottom panel of Figure \ref{MgFe_FeH} \citep[taken from][]{Sales-Silva2024} was used to segregate the high- and low- Mg/Fe populations. 
The blue and red circles in the upper panel of Figure \ref{MgFe_FeH} represent the stars in the low- and high-[Mg/Fe] populations in our sample, respectively.
The stars from globular clusters, and the chemically peculiar, most metal-poor, and Heracles substructure stars discussed above are marked with different symbols in Figure \ref{MgFe_FeH}.

We also detected the presence of other stellar populations in our sample by analyzing other chemical planes. In Figure \ref{multiple_chemical_populations}, we show the [X/Fe]-[Fe/H] planes for Al, K, and Ni, for the high- (shown in red) and low- (shown in blue) Mg/Fe populations. We find evidence for the presence of other populations within the sample of stars with low [Mg/Fe] and [Fe/H] $>-$0.8 (blue distributions). We identify two clumps in the density distribution of the low [Mg/Fe] population (central panels) and multiple peaks in the histograms (right panels) in the chemical planes of the three elements. One of the populations has stars with low [Al/Fe] ratios ([Al/Fe]$< -$0.13), similar to the old and accreted stars found in the Milky Way disk by \citet{Feuillet2022}. We find that these Al-poor stars have mostly eccentric orbits ($<ecc_{Alpoor}>$=0.76$\pm$0.16), which corroborates with their extragalactic nature. The discovery of these stars in the inner region contributes to the mapping and characterization of this accretion event. 

We also identified K-rich stars ([K/Fe]$>$0.25) in the population of low Mg/Fe stars. Unlike C, N, and Na, changes in K abundances are not expected to be due to 
dredge-up, and stellar evolution cannot explain the K enrichment in the red giant stars. \citet{Kemp2018} found 112 K-rich stars mining LAMOST data of $\sim$500,000 Milky Way field stars and hypothesized that a super-Asymptotic Giant Branch binary system may produce this peculiar chemical signature. 
In addition, some stars in globular clusters may exhibit an enrichment in potassium \citep[e.g.,][]{AlvarezGaray2022}, and these stars with high [K/Fe] ratios in the inner Galaxy might be lost stars from globular clusters, similar to the N-rich stars. However, a complication for this scenario is that these stars are not metal-poor, generally presenting [Fe/H]$>$0.0. The chemical peculiarity of K-rich stars may be due to stellar outflows from the super-Asymptotic Giant Branch \citep{Ventura2012}, or pair-instability supernovae \citep{Carretta2013}, which can pollute the stellar atmospheres in dense stellar environments like the bulge. However, \citet{Prantzos2017} found that the enrichment of K in K-rich stars from globular clusters cannot be explained by the material produced by super-AGBs. In fact, the origin of K-rich stars is currently poorly understood. 

In the [Ni/Fe]-[Fe/H] plane of the low-[Mg/Fe] population (shown in the bottom panel of Figure 2), there is a small collection of stars with low-[Ni/Fe] ([Ni/Fe]$<$-0.07), which is much less prominent than those found for [Al/Fe] and [K/Fe]. Ni is an element of the Fe-peak and is mainly ejected into the ISM by type Ia supernovae \citep[e.g.,][]{iwamoto1999,kobayashi2006}. The characteristics of the progenitors (e.g., white dwarf mass or double degenerate white dwarf merger) of these supernovae characterize the yields of these events, including the [Ni/Fe] ratio \citep[e.g.,][]{Seitenzahl2013}. Dwarf satellite galaxies contain Ni-poor stars \citep[e.g.,][]{McWilliam2018,Kirby2019,Montalban2021}. In particular, the Sagittarius Dwarf Galaxy shows Ni-poor stars with metallicity similar to the Ni-poor stars detected here (see Figure 5 in \citealt{Minelli2021}).

In summary, to analyze the chemical abundance gradient, we removed from the low-[Mg/Fe] population the stars that presented some chemical peculiarities (low-[Al/Fe], low-[Ni/Fe], 
or high-[K/Fe]) since such chemical signatures may contaminate the study of chemical gradients. In total, we removed 308, 80, and 535 stars from the low-[Al/Fe], low-[Ni/Fe], and high-[K/Fe] regimes, respectively. It is beyond the scope of this paper to discuss further the nature of these chemically peculiar stars in the inner Milky Way, which we intend to do in a future study.

In addition to chemistry, we also used the plane defined by the orbital parameters $|$Z$_{max}|$ and $ecc$ to characterize the main populations of the bulge-bar, where $|$Z$_{max}|$ is the maximum vertical excursion from the Galactic mid-plane and $ecc$ is the orbital eccentricity ($ecc$ $=$ (R$_{apo}$ $-$ R$_{peri}$)/(R$_{apo}$ $+$ R$_{peri}$), where R$_{peri}$ is the perigalactic distance and R$_{apo}$ is the apogalactic distance). The $|$Z$_{max}|$-$ecc$ plane is a useful tool in the study and  
segregation of stellar populations \citep[e.g.,][]{Eggen1962,Boeche2013,Steinmetz2020,Queiroz2021,Nepal2025}. \citet{Queiroz2021} estimated the orbital parameters (including $|$Z$_{max}|$ and $ecc$) for all stars in our sample, and these are used here
to segregate the stellar populations. Details of the method, including the Galactic potential used to calculate the orbital parameters, are described in section 3 of \citet{Queiroz2021}. The standard error of the orbital parameters of our sample is presented in Figure 2 of \citet{Queiroz2021}. Overall, the eccentricity error is less than 0.02, while the Z$_{max}$ error is less than 0.10 kpc.

In Figure \ref{kinematic_populations}, we plot the density distribution of our sample in the $|$Z$_{max}|$-$ecc$ plane. Using a kinematical analysis, \citet{Queiroz2021} estimated the probability of stars belonging to the Galactic bar; here we assume that stars with a probability greater than 0.5 as being from the bar, or, having bar-driven orbits. In the upper panels of Figure \ref{kinematic_populations}, we show the density distributions for the stars members of the bar with high- (left panel) and low- (right panel) [Mg/Fe] ratios. Both of these bar populations are located mainly near the plane $|$Z$_{max}|\leq$ 1 kpc and they have R$_{Gal}$ $\leq$ 2.5 kpc. In general, bar stars also have orbits with $ecc \gtrapprox$ 0.55, and most of these have very high eccentricity ($ecc \gtrsim$ 0.8), as expected for bar stars.

In the bottom panels of Figure \ref{kinematic_populations}, we show the density distributions in the $|$Z$_{max}|$-$ecc$ plane for the high- and low-[Mg/Fe] stars that are not considered to be part of the bar. 
Clearly, the density distributions for both [Mg/Fe] chemical populations show two denser regions. For low [Mg/Fe] stars, these regions are centered at $|$Z$_{max}|\approx$0.6 kpc and $ecc\approx$0.3 and $|$Z$_{max}|\approx$0.6 kpc and $ecc\approx$0.9. For high [Mg/Fe] stars, the densest regions are centered at $|$Z$_{max}|\approx$1.0 kpc and $ecc\approx$0.32 and $|$Z$_{max}|\approx$0.9 kpc and $ecc\approx$0.9. 
To isolate these main populations, we use limits as indicated by the dashed lines in the figure. 
We limit the sample of low- and high- [Mg/Fe] stars to $|$Z$_{max}|<$ 1.7 kpc and $|$Z$_{max}|<$2.2 kpc, respectively. In addition, we separate stars with low and high eccentricity in both [Mg/Fe] populations. For low-[Mg/Fe] stars, we select stars with $ecc$ lower than 0.45 and higher than 0.7. Meanwhile, the populations of high-[Mg/Fe] stars are selected using $ecc$ lower than 0.50 and higher than 0.75. 

\begin{figure*}
\centering
	\includegraphics[width=13cm]{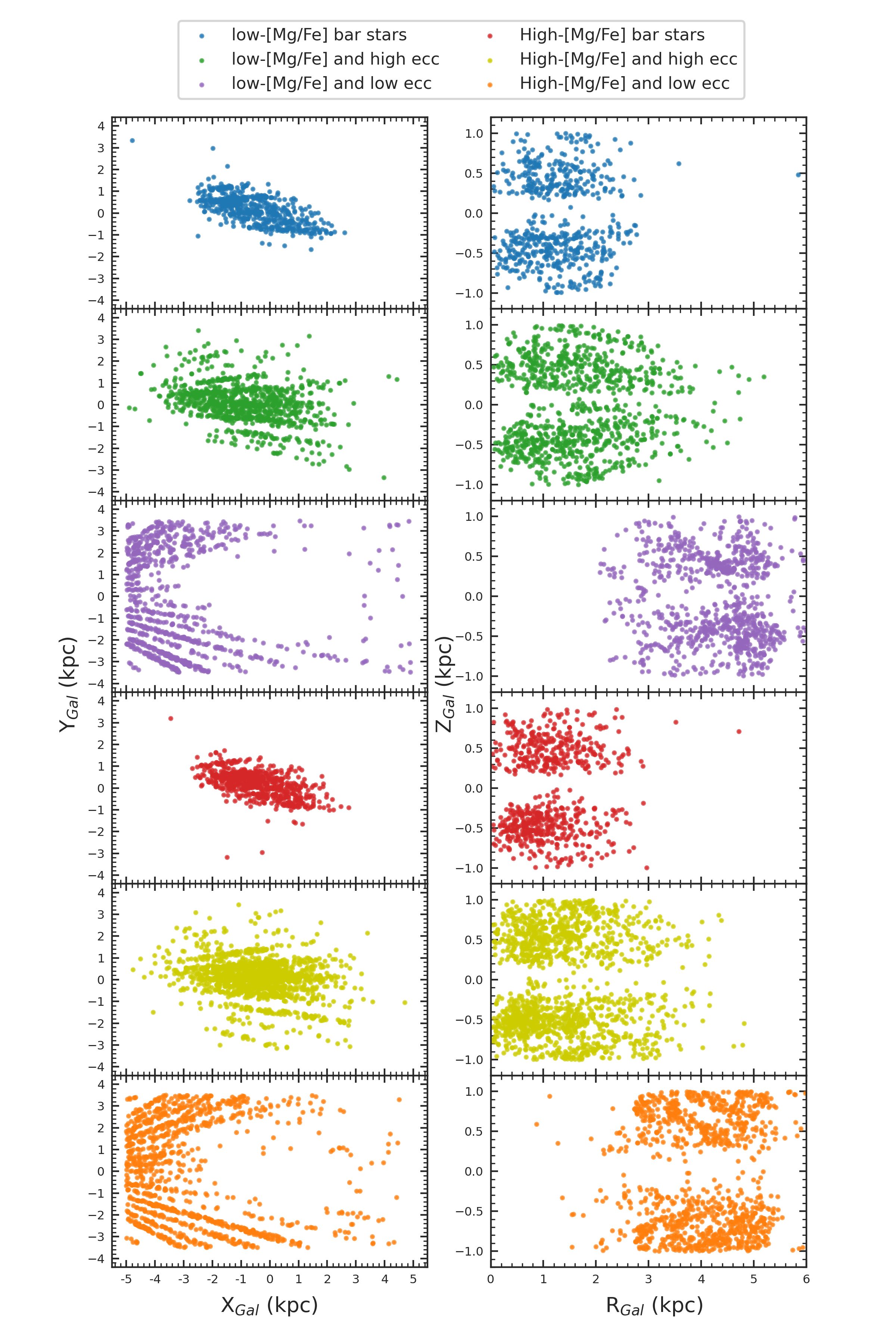}
    \caption{Distribution of six stellar populations present in our sample in the X$_{Gal}$–Y$_{Gal}$ (left panels) and R$_{Gal}$–Z$_{Gal}$ (right panels) planes. Different point colors characterize distinct populations. Blue, green, and purple points represent the low-[Mg/Fe] bar stars, the low-[Mg/Fe] and high $ecc$, and the low-[Mg/Fe] and low $ecc$ populations, respectively, while red, dark yellow and orange represent the high-[Mg/Fe] bar stars, the high-[Mg/Fe] and high $ecc$, and the high-[Mg/Fe] and low $ecc$ populations, respectively.}
    \label{XY_ZR_populations}
\end{figure*}

Summarizing, we chemo-kinematically segregated our inner Galaxy sample into six main populations: two with bar-driven orbits, two with high-eccentricity, but non-bar-driven orbits, and two with low-eccentricity orbits, and each pair composed of stars with low- and high- [Mg/Fe] ratios. Our sample consists of 592 stars with low-[Mg/Fe] from the bar, 698 stars with high-[Mg/Fe] from the bar, 985 stars with low [Mg/Fe] and high $ecc$, non-bar orbits, 1471 stars with high-[Mg/Fe] and high $ecc$, non-bar orbits, 890 stars with low-[Mg/Fe] and low $ecc$, and 1136 stars with high [Mg/Fe] and low $ecc$. In Figure \ref{XY_ZR_populations}, we show the distribution of each population in the X$_{Gal}$-Y$_{Gal}$ and Z$_{Gal}$-R$_{Gal}$ planes. As expected, the bar stars with low- and high-[Mg/Fe] are concentrated near the Galactic center (low R$_{Gal}$). The same is true for populations with high and low [Mg/Fe] and high eccentricity. However, these present a greater spread of distribution in the X$_{Gal}$-Y$_{Gal}$ and Z$_{Gal}$-R$_{Gal}$ planes, probably because they are related to the spheroidal component of the bulge. On the other hand, populations with high and low [Mg/Fe] and low eccentricity are located in the outermost regions of the bulge-bar, generally showing R$_{Gal}>$ 2 kpc.

\section{Abundance gradients of the bulge-bar populations}

\begin{figure*}
\centering
	\includegraphics[width=1.0\textwidth]{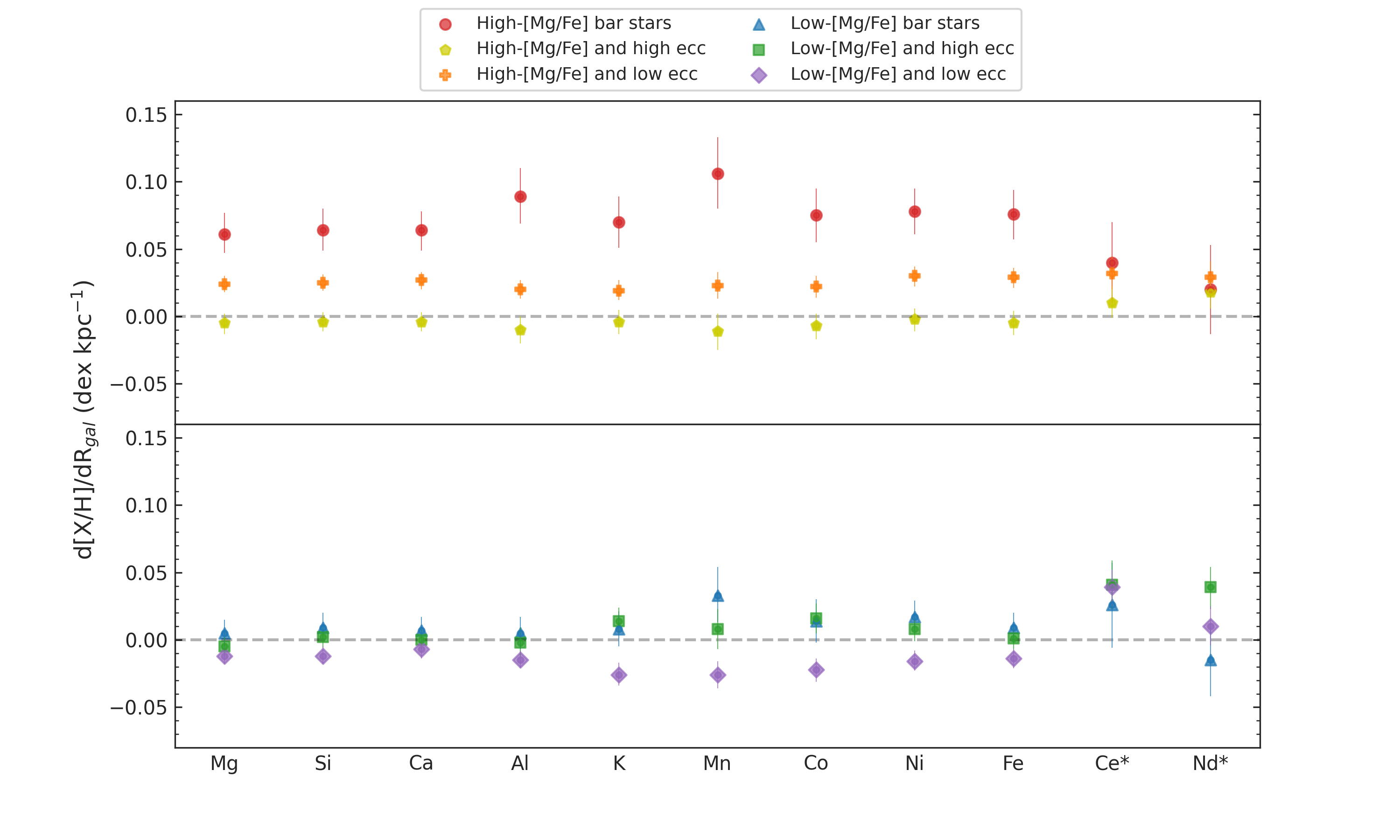}
    \caption{The gradient d[X/H]/dR$_{Gal}$ of Mg, Si, Ca, Al, K, Mn, Co, Ni, Ce, and Nd for different stellar populations. The asterisks at Ce and Nd indicate the use of the smaller sample in the gradient estimation. The legend indicates the stellar populations corresponding to each symbol. The lines in each circle represent the uncertainties of the gradients.
    }
    \label{gradient_mcmc_X_H}
\end{figure*}

\begin{figure*}
\centering
	\includegraphics[width=1.0\textwidth]{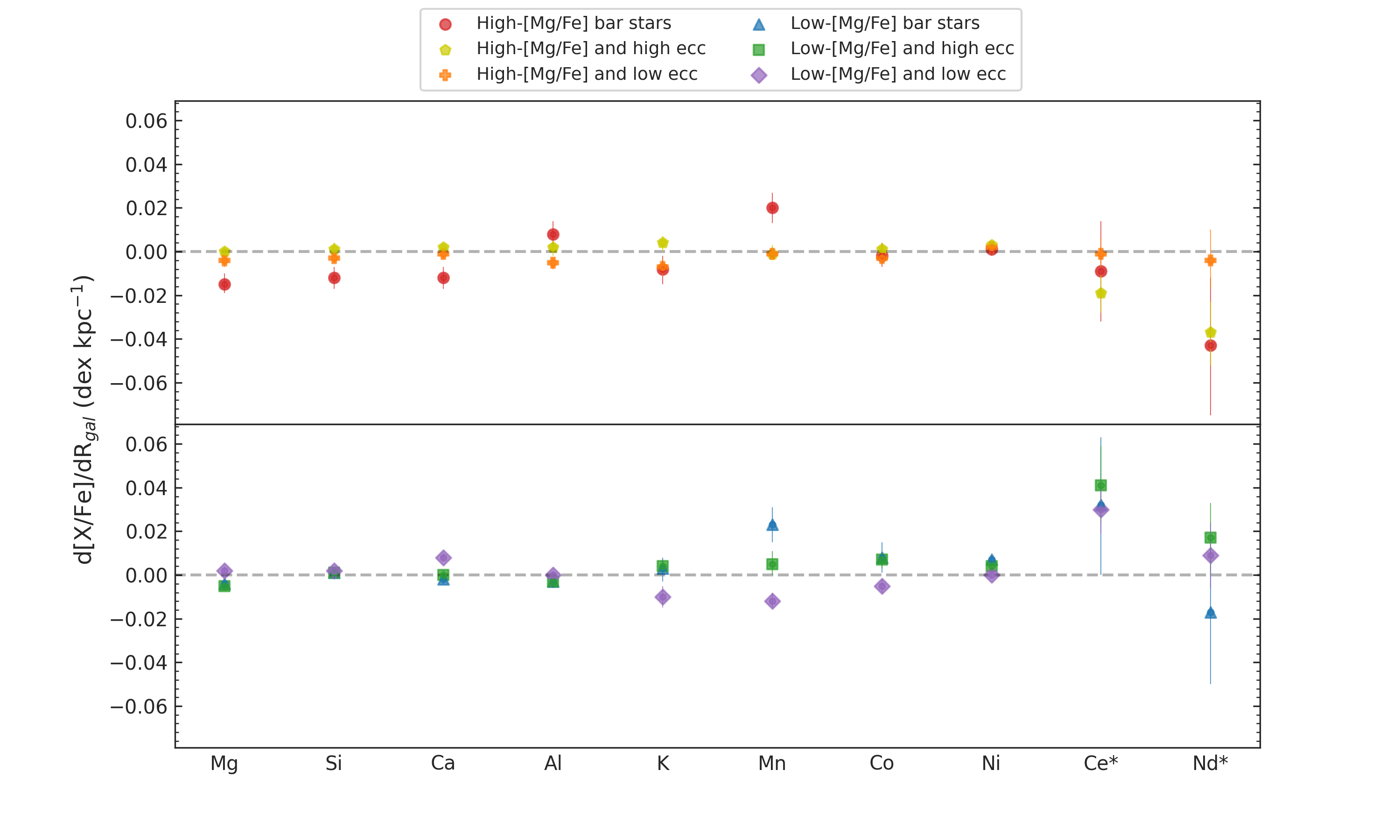}
    \caption{The gradient d[X/Fe]/dR$_{Gal}$ for Mg, Si, Ca, Al, K, Mn, Co, Ni, Ce, and Nd for different stellar populations. The asterisks at Ce and Nd indicate the use of the smaller sample in the gradient estimation. The symbols are as in Fig. \ref{gradient_mcmc_X_H}.
    }
    \label{gradient_mcmc_X_Fe}
\end{figure*}

In this section, we present and discuss the radial abundance gradients of the [X/H] and [X/Fe] ratios for odd-Z (Al and K), alpha (Mg, Si, and Ca), iron-peak (Mn, Co, Ni, and Fe), and neutron-capture (Ce and Nd) elements. We estimate gradients for each of the six main populations discussed in the previous Section. For each elemental gradient, we compute the best linear fits using the maximum likelihood, with associated uncertainties estimated through the Markov-Chain Monte Carlo (MCMC) routine from the {\tt emcee} python package \citep{Foreman-Mackey2013}, as previously done in \citet{Donor2020} and \citet{Sales-Silva2022} for the thin-disk gradients of open clusters. In Tables \ref{low_mgfe_gradients_table} and \ref{high_mgfe_gradients_table} in the Appendix, we show the radial gradients and intercept values estimated for different abundance ratios for the low- and high-[Mg/Fe] populations. Figures \ref{gradient_mcmc_X_H} and \ref{gradient_mcmc_X_Fe} present the best-fit slopes d[X/H]/dR$_{Gal}$ and d[X/Fe]/dR$_{Gal}$ for the elements X = Mg, Si, Ca, Al, K, Mn, Co, Ni, Ce, and Nd derived for the six populations in our sample.

\begin{figure*}
\centering
	\includegraphics[width=13cm]{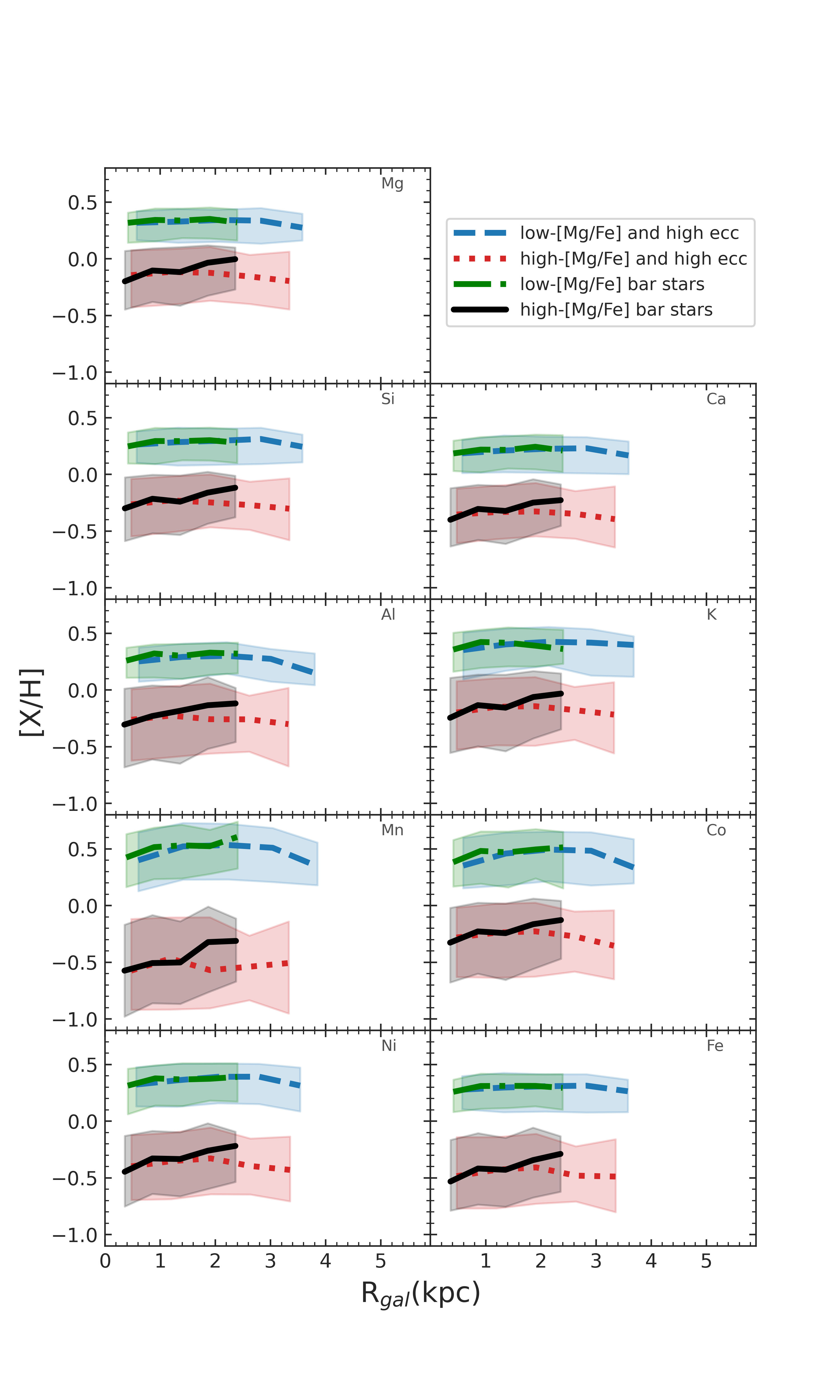}
    \caption{The [X/H] abundance ratio as a function of Galactocentric distance for the high-ecc, non-bar and bar populations. The distinct lines represent the median trends of the bulge-bar populations, with the shaded areas indicating the standard deviation. The legend box above the panels indicates the meaning of the different lines. 
    }
    \label{X_H_vs_rgc_high_ecc}
\end{figure*}

\begin{figure*}
\centering
	\includegraphics[width=13cm]{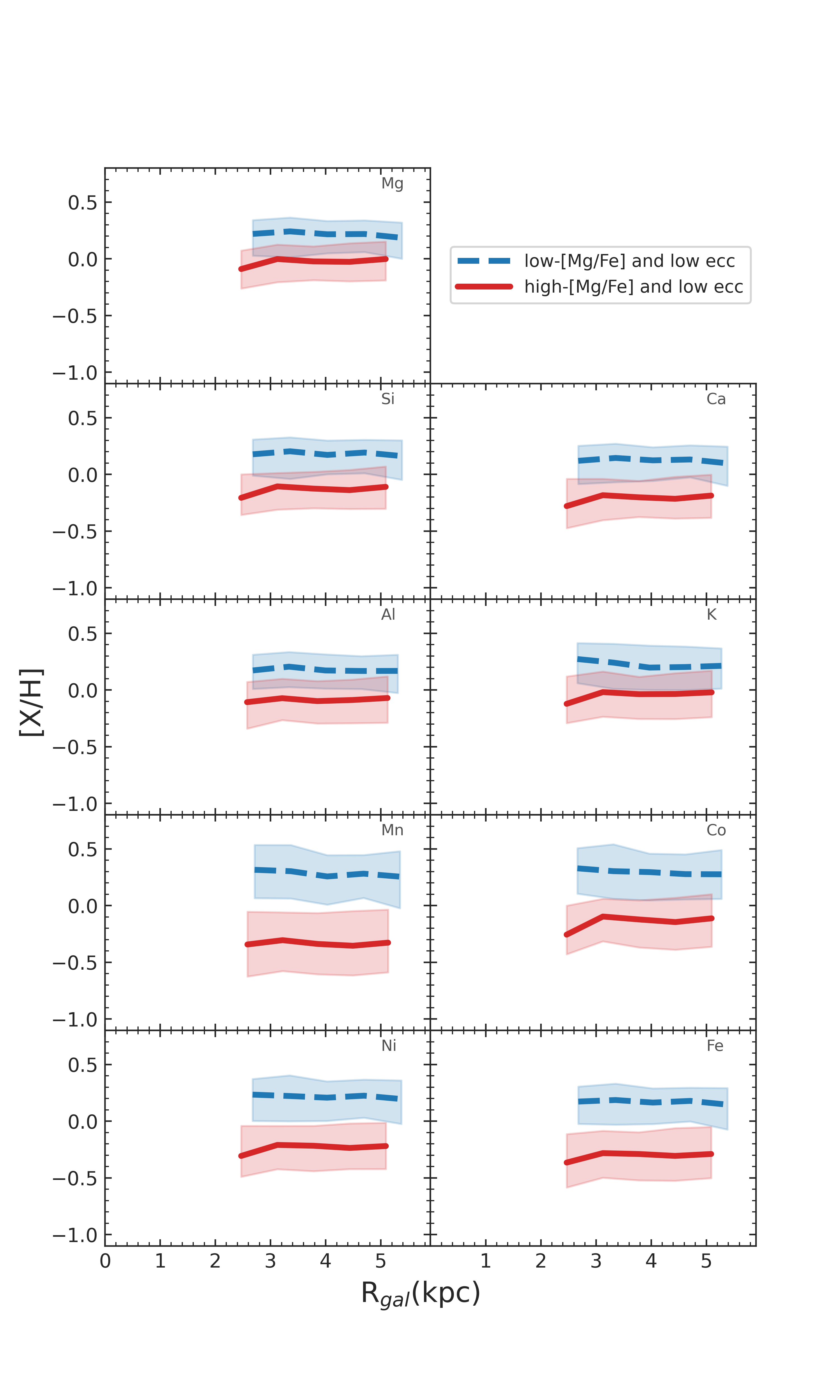}
    \caption{The [X/H] abundance ratio as a function of Galactocentric distance for low-ecc bulge-bar populations. The distinct lines represent the median trends of the bulge-bar populations, with the shaded areas indicating the standard deviation. The legend box above the panels indicates the meaning of the different lines. 
    }
    \label{X_H_vs_rgc_low_ecc}
\end{figure*}

\begin{figure*}
\centering
	\includegraphics[width=13cm]{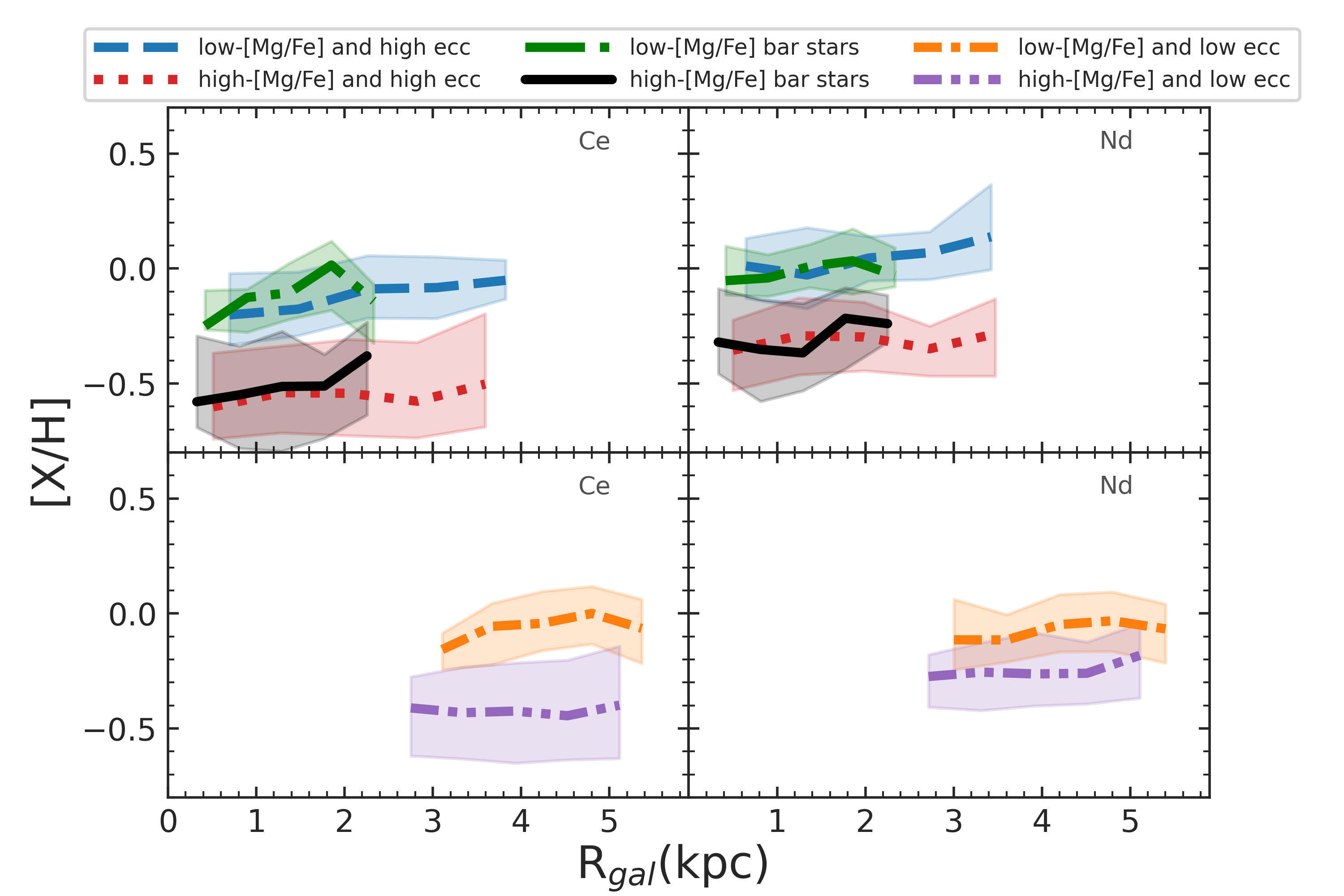}
    \caption{The [Ce/H] and [Nd/H] abundance ratios as a function of Galactocentric distance for bulge-bar populations using abundances from BAWLAS catalog. The distinct lines represent the median trends, with the shaded areas indicating the standard deviation. The legend box above the panels indicates the meaning of the different lines. 
    }
    \label{Ce_Nd_H_vs_rgc}
\end{figure*}

To  visualize the chemical abundance gradients in the different inner galaxy populations studied here better, in Figures \ref{X_H_vs_rgc_high_ecc} and \ref{X_H_vs_rgc_low_ecc}, we show the median [X/H] (dividing R$_{Gal}$ in six bins) as a function of the Galactocentric distance for the elements Si, Mg, Ca, Al, K, Mn, Co, Ni, and Fe in the six inner-Galaxy populations studied here: high-eccentricity, non-bar stars and stars in the bar with high- and low- [Mg/Fe] (Figure \ref{X_H_vs_rgc_high_ecc}), and low-eccentricity stars with high- and low-[Mg/Fe] (Figure \ref{X_H_vs_rgc_low_ecc}). 
In Figure \ref{Ce_Nd_H_vs_rgc}, we present the gradients of the neutron-capture elements for all bulge-bar populations. In Figures \ref{X_Fe_vs_rgc_high_ecc}, \ref{X_Fe_vs_rgc_low_ecc} and \ref{Ce_Nd_Fe_vs_rgc} in the Appendix, we show similar plots, but for the [X/Fe] ratios.

\begin{figure*}
\centering
	\includegraphics[width=15cm]{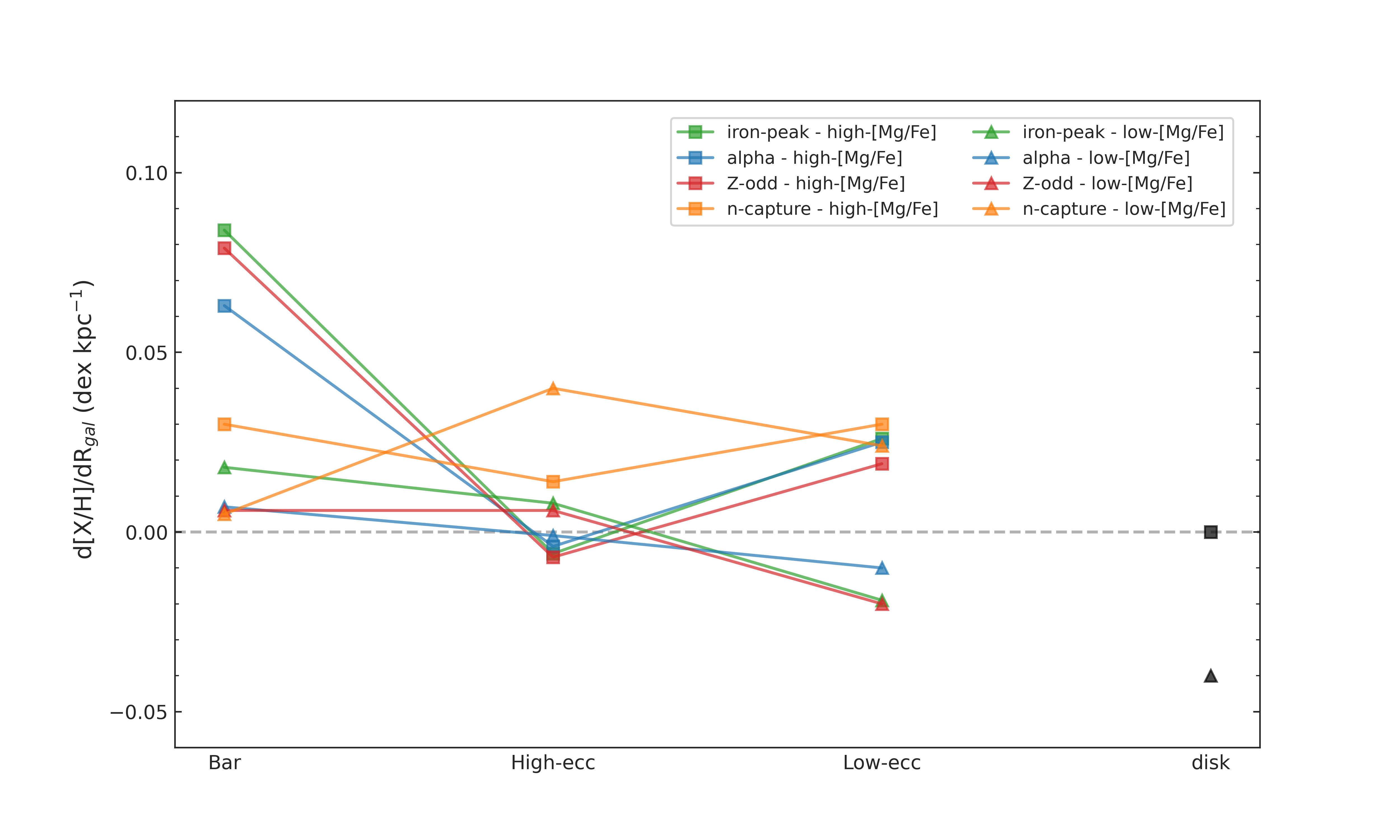}
    \caption{Average gradients of the iron-peak (Mn, Co, Ni, and Fe; green), alpha (Mg, Si, and Ca; blue), odd-Z (Al and K; red), and neutron-capture (Ce and Nd; orange) elements for six stellar populations of the inner galaxy. Triangles and squares represent the low- and high-[Mg/Fe] populations, respectively. We connect the gradients for the same elements with a straight line. We also show the estimated metallicity gradient for the disk using giant field stars at $|Z|<1$kpc \citep[][]{Imig2023} using a square to represent the high-$\alpha$ disk and a triangle for the low-$\alpha$ disk, respectively. 
    }
    \label{summary_gradients_XH}
\end{figure*}

\subsection{Low-[Mg/Fe] populations}

The Milky Way thin disk is a stellar population that follows the low [$\alpha$/Fe] sequence \citep[e.g.,][]{Bensby2003,Fuhrmann2011,Recio-Blanco2014,Hayden2014,Anders2014,Queiroz2020}. The radial metallicity gradient of the thin disk, as measured from open clusters, traces a general decay in metallicity over the range 5 kpc 
$\lessapprox$R$_{Gal}\lessapprox$11-12 kpc, reaching a plateau in the outer disk \citep[e.g.,][]{Spina2022,Magrini2023,Otto2025}, with the slope of the radial metallicity gradient varying between roughly $-$0.05 to $-$0.08 dex kpc$^{-1}$ not counting the plateau region \citep{Joshi2024}.
The disk abundance gradients of [X/Fe] ratios for some elements, such as Al, K, Mg, Si, Ca, Mn, Co, and Ni, may not show a break in the trend \citep[e.g.,][]{Myers2022,Carbajo-Hijarrubia2024}, and, in general, the present [X/Fe] abundance gradients for thin disk populations are approximately zero \citep[e.g.,][]{Spina2021,Myers2022,Magrini2023,Carbajo-Hijarrubia2024}, with some exceptions, such as some elements formed by neutron capture (Ba and Nd, \citealt{Magrini2023}; Ce, \citealt{Myers2022}).

For field stars, metallicity gradients estimated using red giant field stars from the thin disk are also generally within $-$0.02 to $-$0.07 dex kpc$^{-1}$ for $|$Z$|<$1.75 kpc \citep[e.g.,][]{Boeche2014,Anders2017,Eilers2022,Willett2023,Anders2023,Imig2023,Johnson2025}. In particular, using a red giant and red clump sample from APOGEE,  \citet{Imig2023} found that the metallicity gradient flattens with distance from the midplane. 
Using a representative sample of red giant stars from the GALAH survey, \citet{Wang2024} estimated a metallicity gradient of $\sim$  $-0.029\pm0.003$ dex kpc$^{-1}$  for the disk in the solar neighborhood (R$_{Gal}$=7-9 kpc), a gradient similar to that found by \citet{Stanghellini2018} for thin disk planetary nebulae (d[O/H]/dR$_{Gal} $=$ -$0.021) over 
a wider range of R$_{Gal}$ (6-16 kpc).

In addition, chemical gradients evolve over time \citep[e.g.][]{Minchev2018,Ratcliffe2023,Anders2017,Willett2023,Anders2023}. Using combined asteroseismic data of red giant stars from CoRoT and spectroscopic observations from APOGEE, \citet{Anders2017} found that the metallicity gradient flattens from a negative value of $-$0.066 dex/kpc at ages of 1-4 Gyr to $-$0.030 dex/kpc for stars with ages between 6 and 10 Gyrs. Similar results were also found in other studies (e.g., \citealt{Willett2023}, using asteroseismic ages from K2 and APOGEE data; \citealt{Anders2023}, with APOGEE and Kepler data, and machine learning to transfer labels from the seismic ages of \citealt{Miglio2021}).
Using a red giant sample from APOGEE and estimated birth radii, \citet{Ratcliffe2023} found flattening and inversion of the chemical gradient of the disk in older populations due to radial migration. In the latter study, the [X/H] gradients for Al, Mg, Ca, and Mn, which are initially negative, tend to flatten over time. In the case of [X/Fe] ratios, the gradient evolves from positive to approximately zero or slightly negative values for Al, Mg, Ca, and Mn.

Given observational limitations, few studies in the literature have analyzed abundance gradients with a focus on the most central region of the Milky Way Galaxy. 
\citet{Imig2023} found a negative metallicity gradient for low-$\alpha$ disk field stars located roughly between 2-5 kpc from the Galactic center.  
In addition, \citet{Imig2023} found a slightly positive metallicity gradient for the inner high-$\alpha$ disk.

In our sample, two of the low-[Mg/Fe] populations show an approximately flat metallicity gradient (close to the zero slope dashed line in the lower panel of Figure \ref{gradient_mcmc_X_H}): the low [Mg/Fe] members of the bar and the low [Mg/Fe] high eccentricity stars.
In contrast, the low-[Mg/Fe] population of stars with low eccentricity orbits is the only one that exhibits a slight negative gradient (d[Fe/H]/dR$_{Gal} $=$ -$0.014$\pm$0.007). The latter stars are located at R$_{Gal}>$2.0 kpc (i.e., are not in the regions closest to the Galactic center, like the other low-[Mg/Fe] populations), and they exhibit more circular orbits when compared to those in the other two low-[Mg/Fe] populations, being the population in our sample which is more similar to the thin disk.  
It is important to note, however, that the [Fe/H] gradient of the low-[Mg/Fe] low eccentricity population is significantly less steep than the gradient of the outer regions of the thin disk, indicating that there may be a break in the gradient trend of the thin disk at R$_{Gal}\approx$ 5-6 kpc, if we compare, for example, with the APOGEE gradients $-$0.06 dex/kpc of field stars with distances larger than $\sim$5 kpc \cite[][]{Imig2023}. Using APOGEE DR16 red giant stars covering galactocentric radii of 0$<$R$_{Gal}<$20 kpc, \citet{Eilers2022} did not find this break in the gradient of the inner low-$\alpha$ disk.

For the population of low-[Mg/Fe] low-eccentricity stars, the mean [Fe/H] abundance at R$_{Gal}$ $\sim$ 2 kpc is $<$[Fe/H]$>$ $\sim$+0.15 (Figure \ref{X_H_vs_rgc_low_ecc}), which is similar to the mean metallicity of $<$[Fe/H]$>$ $\sim$ +0.12 found for a sample of stars within 30 pc of the Galactic center in the early near-infrared high-resolution study of \citep[][]{Cunha2007} (see also more recent work of \citealt{Nandakumar2025} and \citealt{Ryde2025}). Such results may point to the constancy of metallicity towards the center of the Galaxy for the giant stars.

In general, the [X/H] gradients of the other elements for the low-[Mg/Fe] populations follow the trend shown by [Fe/H], except for the neutron-capture elements cerium and neodymium (see Figure \ref{gradient_mcmc_X_H}). Both Ce and Nd can be produced through either slow or rapid neutron captures (the s-process and the r-process, respectively), with Ce being mostly an s-process element and Nd having a more significant fraction coming from the r-process \citep{Prantzos2020}. The s-process synthesizes Ce and Nd mainly in low- and intermediate-mass AGB stars \citep{Lugaro2003}, while the r-process can produce these elements in neutron star{-neutron star} \citep[e.g.,][]{Watson2019,Thielemann2017} or black hole–neutron star \citep[e.g.,][]{Ekanger2023} mergers as well as core-collapse \citep[e.g.,][]{Wheeler1998,Ekanger2023}, or magneto-rotational  \citep[e.g.,][]{winteler2012,Reichert2021,Reichert2023} supernovae. As pointed out in \citet{Sales-Silva2024}, the low [Ce/Nd] ratios observed in (mainly metal-poor) bulge-bar stars indicate ancient enrichment dominated by the r-process, possibly a signature of magneto-rotational supernovae yields generated through massive stars. \citet[][]{Razera2022} also note that the chemical abundances of these heavy elements in the metal-poor bulge population may have a significant contribution from massive rotating stars. The Ce and Nd gradients for stars with low [Mg/Fe] and low $ecc$ show positive values (mainly Ce), but these also present larger uncertainties, associated with the larger abundance errors estimated in \citet{Hayes2022}. 

The [X/Fe] gradients indicate the production of a given element relative to the production of iron at different R$_{Gal}$, unlike the [X/H] ratios, which have a strong dependence on star formation.  
[X/Fe] gradient values close to zero indicate a similar trend with R$_{Gal}$ between the element and iron with Galactocentric distance. Most of the [X/Fe] gradients for the different elements for low-[Mg/Fe] populations show a constant trend, with approximately zero gradient. The exceptions are: Ce and Nd for the three low-[Mg/Fe] populations, Mn for the low-[Mg/Fe] bar members, and for low-[Mg/Fe] low-$ecc$ population, and K for low-[Mg/Fe] low-$ecc$ population. 
In the thin-disk context, near-zero [X/Fe] slopes are also found for the low-$\alpha$ population \citep[e.g.,][]{Ratcliffe2023b}. For the [Ce/Fe] gradient, \citet{Ratcliffe2023b} found a reversal behavior depending on the bin of ([Fe/H], [Mg/Fe]). 
\citet{Casali2023} found a strong dependence of the [Ce/Fe] abundances on R$_{Gal}$, and related this behavior of [Ce/Fe] to different star formation histories and the metallicity dependence of stellar yields. Corroborating this result, \citet{Sales-Silva2024} found a spatial chemical dependence of the Ce and Nd abundances in the inner Galaxy.

Overall, manganese is the iron-peak element that displays the largest difference between the gradients of the different low-[Mg/Fe] populations. Mn is a pristine tracer of Type Ia supernovae \citep{iwamoto1999,kobayashi2006}, more than iron, which exhibits significant production by core-collapse supernovae \citep[e.g.,][]{Rodriguez2023}. Among the elements in the Fe peak, the [Mn/H] gradient for the low-[Mg/Fe] and low eccentricity population is steeper 
(d[Mn/H]/dR$_{Gal} $=$ -$$0.026^{+0.010}_{-0.010}$). On the other hand, the low-[Mg/Fe] of the bar population presents a positive gradient (d[Mn/H]/dR$_{Gal}$=0.033) but with a significant uncertainty ($\approx$0.020).

The [K/H] gradient has the same negative value as the [Mn/H] gradient. Like Fe, this odd-Z element can be ejected into the ISM by Type Ia and Type II supernovae, and its production depends on several parameters, such as the metallicity of the Type Ia supernova progenitor, stellar rotation \citep{Prantzos2018}, and the neutrino processes in core-collapse supernovae \citep{Kobayashi2011}. Potassium is also produced in super-Asymptotic Giant Branch \citep{Ventura2012} and pair-instability supernovae \citep{Carretta2013}.

The [X/H] gradients of the alpha elements are more similar to the Fe gradient than the Mn gradient. Alpha elements (such as Mg, Ca, and Si) are the main tracers of core-collapse supernovae \citep{woosley-weaver1995}. Overall, the gradients of these elements show approximately flat values for the three low-[Mg/Fe] populations, with the low-eccentricity population showing slightly negative values.

\subsection{High-[Mg/Fe] populations}

In the upper panels of Figures \ref{gradient_mcmc_X_H} and \ref{gradient_mcmc_X_Fe}, we show the gradients for the high-[Mg/Fe] populations and these results can be compared to those of the low-[Mg/Fe] populations discussed in the previous subsection and shown in the bottom panels of Figures  \ref{gradient_mcmc_X_H} and \ref{gradient_mcmc_X_Fe}. 

For most of the elements, the [X/H] gradients for high [Mg/Fe] high-eccentricity stars have slopes close to zero, and this is also what we find for the gradients of the low [Mg/Fe] high eccentricity population. The only exceptions are the neutron-capture elements Ce and Nd, which show positive gradients. On the other hand, for bar populations and stars with low $ecc$, there is a significant difference in the [X/H] gradient between low- and high-[Mg/Fe] stars. In general, the low- and high-[Mg/Fe] populations with low $ecc$ show opposite signs of the [X/H] gradient, being positive for high-[Mg/Fe] stars and negative for low-[Mg/Fe] populations. The exceptions again are the neutron capture elements Ce and Nd, which show the same (positive) gradients between the low- and high-[Mg/Fe] stars. For example, the metallicity gradient for the high-[Mg/Fe] low $ecc$ stars is d[Fe/H]/dR$_{Gal}$ = +$0.029^{+0.007}_{-0.008}$ dex kpc$^{-1}$, while the metallicity gradient for the low-[Mg/Fe] low $ecc$ population is d[Fe/H]/dR$_{Gal}$ = $-$0.014$\pm$0.007 dex kpc$^{-1}$. The high-[Mg/Fe] low $ecc$ stars present more circular orbits compared to the stars of the other high-[Mg/Fe] populations, and they are located at R$_{Gal} > $2.0 kpc. Such a population resembles both chemically (high-[Mg/Fe]) and kinematically the thick disk stars. As shown by observations \citep[e.g.,][]{Carrell2012,Sun2024} and chemodynamical simulations for a structural thick disk \citep[e.g.,][]{Minchev2014,Miranda2016,Li2017}, the metallicity gradient of the thick disk may present positive and inverted values compared to the thin disk, a consequence of the flaring of mono-age populations, which causes a mixture of stars of different ages at a given Z, an effect known as Simpson's paradox or Yule-Simpson effect \citep[][]{Minchev2019}.
In addition, the thick disk had a turbulent formation and radial migration of its old population, which influences the characterization of its chemical gradient. 
Thus, opposite trends in the gradients between the low- and high-[Mg/Fe] populations with low $ecc$ may correspond to different epochs (or mixtures of stellar population ages) and formation processes of these populations, as for stars in thin and thick disks.

The stellar populations in our sample that present steeper positive [X/H] gradients (d[X/H]/dR$_{Gal} > $+0.060 dex kpc$^{-1}$, except for Ce and Nd) are the high-[Mg/Fe] stars of the bar. In particular, the element with the steepest gradient is [Mn/H], with a slope of d[Mn/H]/dR$_{Gal}$=+0.106 dex kpc$^{-1}$, which indicates increasing Mn abundances with increasing R$_{Gal}$. When ratioed to Fe, we observe that Mn is the only element that has a positive value for the [X/Fe] gradient (d[Mn/Fe]/dR$_{Gal}$ = +0.023 dex kpc$^{-1}$) in the high-[Mg/Fe] stars of the bar. Within the uncertainties, the other elements have [X/Fe] gradients that are nearly zero with only a small scatter in the mean (d$<[X/Fe]>$/dR$_{\rm Gal}$ = -0.007$\pm$0.010 dex kpc$^{-1}$, considering all elements other than Mn, Ce, and Nd). 
The steep and positive [X/H] gradients found for the high-[Mg/Fe] bar population are possibly related to the formation and evolution of bar characteristics and not due to differences in the nucleosynthesis of the elements, given the zero gradients of the [X/Fe] ratios.

The presence of a bar is a relatively common feature in disk galaxies ($\sim$50-70\% of disk galaxies in the local universe) \citep{Menendez-Delmestre2007,Nair2010,Gavazzi2015,Lee2019}, suggesting that its formation is a fundamental process in galaxy evolution. The role of different mechanisms and the precise initial conditions for bar formation remain under discussion \citep[e.g.,][]{Rosas-Guevara2025}. The dominant theory for bar formation is disk instabilities \citep[e.g.,][]{Athanassoula2002,Debattista2006}. However, other processes (such as gravitational interactions) may affect the evolution of a bar. Observable properties, such as chemical gradients, help reveal the history of the bulge-bar. 
Such steep positive gradients found for all elements in the high-[Mg/Fe] bar stars may be reminiscent of such gradients in high-redshift (z $\sim$ 4-10) galaxies reported by \citet[]{Tripodi2024} based on JWST spectra. This particular redshift interval probed by \citet[]{Tripodi2024} corresponds to look-back times of about 12-13 Gyr.

Our bar sample extends to $\approx2.5$ kpc and includes only a few stars near the midplane ($|Z|<0.2$ kpc; see Figure \ref{XY_ZR_populations}). Thus, it primarily traces the upper part of the peanut-shaped bulge rather than the full midplane bar, which is thought to extend to $\approx$4-5.5 kpc \citep[e.g.,][]{hilmi20,bland-hawthorn16}. In this sense, our stars occupy the same region as the red triangles in Figure 13 of \citet{Fragkoudi2018}, which show a positive metallicity gradient along the bar’s major axis up to 2.5-3 kpc, followed by an inversion beyond that radius. This pattern is also evident in our data (Figures \ref{gradient_mcmc_X_H} and \ref{X_H_vs_rgc_high_ecc}).

The positive [Fe/H] gradient observed among our high-[Mg/Fe] bar stars (Figure \ref{X_H_vs_rgc_high_ecc}) probably reflects an age variation along the peanut structure: stars closer to the Galactic center are older, while those toward the bar end are younger by a few Gyr. This interpretation agrees with the N-body simulations of \citet{DiMatteo2019}, whose Figures 5 and 6 show that, indeed, stars at larger bar radii (and away from the midplane) are both younger and more metal-rich.

The convergence of [Fe/H] between the high- and low-[Mg/Fe] bar populations near $R_{\mathrm{Gal}}\approx2.5$ kpc may therefore indicate that the high-[Mg/Fe] stars become progressively younger and more metal-rich with increasing radius. This inverse age-metallicity trend within the high-[Mg/Fe] sequence is also consistent with Figure \ref{MgFe_FeH}, where [Mg/Fe], and thus stellar age, decreases by $\approx0.25$ dex as [Fe/H] increases from $-$1 to 0.

Summarizing our results, Figure \ref{summary_gradients_XH} presents the average gradients for the iron peak, alpha, Z-odd, and neutron capture elements for the different populations of the inner Galaxy, and for the metallicity of the low- and high-$\alpha$ disks of \citet{Imig2023} that used only abundances of APOGEE DR17 (without empirical corrections) and separated the disk populations with boundaries in the [Mg/Fe]-[Fe/H] plane similar to our limits. We observe that the estimated gradient for the Fe-peak elements for the high-[Mg/Fe] low eccentricity stars is steeper than the estimated metallicity gradient for the high-$\alpha$ disk at $|Z|<1.0$kpc 0.000$\pm$0.005 dex kpc$^{-1}$ \citep{Imig2023}. Some studies of the thick disk \citep[e.g,][]{Anders2014,Hayden2014,Wang2024,Chen2025} show similar metallicity gradient values than the gradient of the Fe-peak elements for the high-[Mg/Fe] low eccentricity stars. Additionally, we detect a significant difference between the Fe-peak element gradients for the low-[Mg/Fe], low eccentricity stars and the metallicity estimate for the low-$\alpha$ disk at $|Z|<1.0$kpc $-$0.040$\pm$0.002 dex kpc$^{-1}$ \citep{Imig2023}. 

Clearly, in Figure \ref{summary_gradients_XH}, the elements produced by the neutron capture (orange symbols) generally exhibit a different behavior compared to the other elements in various populations of the inner region. This behavior for these elements is also observed on the disk \citep[e.g,][]{Myers2022,Magrini2023}. Understanding evolution and neutron capture element distribution is complex due, for example, to its strong and complicated dependence on [Fe/H] and age, as shown in studies of thin disk populations \citep[e.g,][]{Sales-Silva2022,Casali2023}. Thus, beyond the significant uncertainties in the gradient estimates, the distinct gradient of Ce and Nd should be strongly associated with the age and [Fe/H] of the populations in the inner region of the Galaxy. The production efficiency of heavy s-process elements such as Ce and Nd depends strongly on metallicity, since a distinct iron content causes a change in neutrons available per seed (iron) nucleus used by the s-process, while the nature of the relationship between heavy elements and age is complex and not fully understood \citep[e.g,][]{Baratella2021}. The enrichment of heavy elements in the interstellar medium is delayed by the slow evolution of AGB stars, creating a complex link to stellar age, while uncertainties in r-process sources add further complication.

\section{Conclusions}

We present radial abundance gradients for $\sim$8000 stars from the bulge-bar selected by \citet{Queiroz2021}, using chemical abundances from APOGEE DR17 for the elements Mg, Si, Ca, Al, K, Mn, Co, Ni, and Fe. In addition, we analyzed the gradients of the heavy elements Ce and Nd using abundances from the DR17 BAWLAS catalog \citep{Hayes2022} for a subsample of approximately 2000 bulge-bar stars. This study includes analyses of $\alpha$, odd-Z, iron-peak, and neutron-capture elements, providing a chemically broad view of the chemical abundance gradients in the inner Milky Way.

The bulge-bar is known to be composed of a complex mixture of multiple populations. 
Following \citet{Sales-Silva2024} we removed from our bulge-bar sample 37 chemically peculiar stars (mostly N-rich), 37 probable globular cluster members, and 33 possibly ex-situ stars from the Heracles substructure. 
We segregated our sample into low- and high-[Mg/Fe] stars using the [Mg/Fe]-[Fe/H] plane and found two chemical populations within the low-[Mg/Fe] population: one Al-poor and one K-rich. In addition, there was a small group of Ni-poor stars. 
These Al-poor, K-rich, and Ni-poor stars were removed from the low-[Mg/Fe] population sample.

To perform a detailed analysis of the chemical radial gradients of the different kinematical populations in the bulge-bar sample, we further segregated low- and high-[Mg/Fe] populations using the  $|$Z$_{max}|$-$ecc$ orbital plane into six main populations: two with bar-driven orbits, two with eccentric orbits, and two with low-eccentricity orbits, with each pair being composed of stars with low and high [Mg/Fe].

Our results can be summarized as follows:

\begin{itemize}
\item[-] The low-[Mg/Fe] stars with low-eccentricity orbits located at radii 2.0 $<$ R$_{Gal}$ $<$ 6.0 kpc show a slightly negative value of the metallicity gradient (d[Fe/H]/dR$_{\rm Gal}$ = $-$0.014$\pm$0.007 dex kpc$^{-1}$). This population presents orbital and chemical characteristics similar to those of thin disk stars. However, thin disk stars at Galactocentric distances larger than $\sim$5.0-6.0 kpc show a much steeper metallicity gradient of $-$0.06 dex kpc$^{-1}$ \citep[][]{Imig2023}. 
The change in slope at around $\sim$ 5 - 6 kpc may point to a break in the global [Fe/H] gradient trend of the thin disk as it transitions to flatter gradients in the inner regions of the Galaxy.

\item[-] The other two low-[Mg/Fe] populations in our sample show approximately flat profiles for the metallicity gradients. For the low [Mg/Fe] high eccentricity population, d[Fe/H]/dR$_{\rm Gal}$ = 0.001$^{+0.09}_{-0.008}$ dex kpc$^{-1}$, while for the low [Mg/Fe] bar members, d[Fe/H]/dR$_{\rm Gal}$ = $+$0.009$^{+0.011}_{-0.011}$ dex kpc$^{-1}$.

\item[-] In general, the [X/H] gradients of the other elements in the low-[Mg/Fe] populations follow similar trends to those followed by Fe, with the exception of those elements produced by neutron captures, Ce and Nd. The Ce and Nd gradients for stars with low [Mg/Fe] and low $ecc$ show positive slopes, mainly for Ce, although the latter have larger uncertainties. Ce and Nd present a strong dependence of their production on age and [Fe/H], which probably accounts for the distinct gradient of Ce and Nd compared to other elements.

\item[-] The [X/Fe] gradients for all elements of the three low-[Mg/Fe] populations are approximately zero, indicating similar trends between those elements and iron with Galactocentric distance. The exceptions are Ce for the three low-[Mg/Fe] populations (d$<$[Ce/Fe]$>$/dR$_{\rm Gal}$ = $+$0.034$\pm$0.006 dex kpc$^{-1}$), Mn for the low-[Mg/Fe] from the bar (d[Mn/Fe]/dR$_{\rm Gal}$ = 0.023$^{+0.008}_{-0.008}$ dex kpc$^{-1}$) and low-[Mg/Fe] populations with low $ecc$ (d[Mn/Fe]/dR$_{\rm Gal}$ = $-$0.012$^{+0.003}_{-0.004}$ dex kpc$^{-1}$), and K for the low-[Mg/Fe] and low $ecc$ population (d[K/Fe]/dR$_{\rm Gal}$ = $-$0.010$^{+0.005}_{-0.005}$ dex kpc$^{-1}$). (All of these slopes are shown in Table \ref{low_mgfe_gradients_table}).

\item[-] In general, the [X/H] gradients for both the low- and high-[Mg/Fe] populations having orbits with high eccentricity present approximately constant, near-zero slopes. In contrast, the low- and high-[Mg/Fe] populations having low $ecc$ orbits have [X/H] gradients with slopes of opposite signs, being positive for high-[Mg/Fe] stars and negative for low-[Mg/Fe] populations. In terms of Galactocentric distance, the high-[Mg/Fe], low $ecc$ population located at 2.0~$<$~R$_{Gal}~<$~6.0 kpc chemically and kinematically resembles the thick disk stars. This similarity also extends to the [Fe/H] gradient, since the thick disk can show positive gradient values \citep[e.g.,][]{Carrell2012,Sun2024}.

\item[-] We found that the two bar populations (low- and high-[Mg/Fe]) exhibit distinct radial gradients with the high-[Mg/Fe] showing a positive gradient much steeper than that for the low-[Mg/Fe] bar stars, which are overall flat. This difference in the radial gradients for the bar populations may be linked to the formation of the low- and high-[Mg/Fe] components at different times and via distinct Galactic evolutionary scenarios, as occurred in the formation of the thin and thick disk. The bulge formation simulations \citep[e.g.,][]{Debattista2017} and the chemical similarity of the thin disk stars to the low-[Mg/Fe] bar stars are in line with the formation of this bar population through secular evolution from an early low-[Mg/Fe] thin disk. Recently, through dynamical and chemical analysis, \citet{Pandey2025} obtained results indicating that the high-[Mg/Fe] bar stars originated via secular evolution from a high-$\alpha$ disk.
The approximately constant value of the [X/Fe] gradients for high-[Mg/Fe] bar stars indicates that the steep value of the [X/H] gradient is possibly not to issues of nucleosynthesis of the elements. We found a convergence of [Fe/H] between the high- and low-[Mg/Fe] bar populations near $R_{\mathrm{Gal}}\approx2.5$ kpc, indicating that the high-[Mg/Fe] stars become progressively younger and more metal-rich with increasing radius.

\item[-] We also found that the gradients for the high-[Mg/Fe] bar stars are significantly different than those of the other high-[Mg/Fe] bulge populations, with the bar population showing more positive and steep values for all elements. Our bar sample extends to $\approx2.5$ kpc and includes only a few stars near the midplane ($|Z|<0.2$ kpc). So, the steep positive [X/H] gradients observed among our high-[Mg/Fe] bar stars may also reflect an age variation along the upper part of the peanut-shaped bulge: stars closer to the Galactic center are older, while those toward the bar end are younger by a few Gyr. This interpretation agrees with the N-body simulations.
If the high-[Mg/Fe] bar population represents the oldest stars, these results may also be related to the steep positive metallicity gradients derived by \citet{Tripodi2024} for high-redshift, z $\sim$ 4-10 galaxies observed with JWST.
Unfortunately, bulge-bar stars suffer from inaccurate age estimates, making a detailed analysis of the temporal evolution of radial gradients challenging at this point. The lack of accurate ages is a limiting factor in archeological studies of the inner Galactic region. Future space missions providing high-precision asteroseismology in dense fields of the inner Galaxy \citep[][]{Miglio2021b} or more robust method for estimating age through machine learning may contribute to a better understanding of the inner Milky Way gradients.

\end{itemize}

Our results provide important observational constraints for Galactic models, particularly for bulge-bar evolution, contributing to the determination of the influence of processes such as interactions with dark matter halo and gas physics related to star formation and stellar feedback.


\begin{acknowledgments}

JVSS acknowledges the PCI programme under grant 313980/2025-0 and 300523/2026-2. S.D. acknowledges CNPq/MCTI for grant 306859/2022-0 and FAPERJ for grant 210.688/2024. D.S. thanks the National Council for Scientific and Technological Development – CNPq.
V.L.T. acknowledges support from the CNPq through the Postdoctoral Junior (PDJ) fellowship, process No. 152242/2024-4.
TM acknowledges support from MCIN for the project\textit{PLAtoSOnG} from its grant PID2023-146453NB-100, PI: Beck). PMF acknowledge support for this research from the National Science Foundation (AST-2206541). IM acknowledges support by the Deutsche Forschungsgemeinschaft under the grant MI 2009/2-1.
J.G.F-T gratefully acknowledges the support provided by ANID Fondecyt Regular No. 1260371, ANID Fondecyt Postdoc No. 3230001 (Sponsoring researcher), the Joint Committee ESO-Government of Chile under the agreement 2023 ORP 062/2023 and the support of the Doctoral Program in Artificial Intelligence, DISC-UCN.
MZ acknowledges funding from ANID BASAL Center for Astrophysics and Associated Technologies (CATA) FB210003 and from FONDECYT Regular grant No. 1230731.
Funding for the Sloan Digital Sky Survey IV has been provided by the Alfred P. Sloan Foundation, the U.S. Department of Energy Office of Science, and the Participating Institutions. SDSS-IV acknowledges support and resources from the Center for High-Performance Computing at the University of Utah. The SDSS website is www.sdss.org.

SDSS-IV is managed by the Astrophysical Research consortium for the Participating Institutions of the SDSS Collaboration including the Brazilian Participation Group, the Carnegie Institution for Science, Carnegie Mellon University, the Chilean Participation Group, the French Participation Group, HarvardSmithsonian Center for Astrophysics, Instituto de Astrofísica de Canarias, The Johns Hopkins University, Kavli Institute for the Physics and Mathematics of the Universe (IPMU)/University of Tokyo, Lawrence Berkeley National Laboratory, Leibniz Institut fur Astrophysik Potsdam (AIP), Max-Planck-Institut fur Astronomie (MPIA Heidelberg), Max-Planck Institut fur Astrophysik (MPA Garching), Max-Planck-Institut fur Extraterrestrische Physik (MPE), National Astronomical Observatory of China, New Mexico State University, New York University, University of Notre Dame, Observatório Nacional/MCTI, The Ohio State University, Pennsylvania State University, Shanghai Astronomical Observatory, United Kingdom Participation Group, Universidad Nacional Autónoma de México, University of Arizona, University of Colorado Boulder, University of Oxford, University of Portsmouth, University of Utah, University of Virginia, University of Washington, University of Wisconsin, Vanderbilt University, and Yale University.

\end{acknowledgments}





%

\vspace{5mm}
\facilities{Sloan (APOGEE)}


\software{matplotlib \citep{Hunter2007}, Numpy \citep{Harris2020}, Scipy \citep{Virtanen2020}}



\appendix

\section{Radial abundance gradients and intercept values}

\begin{table*}[h]
	\centering
	\scriptsize
	\caption{Radial abundance gradients (dex kpc$^{-1}$) and intercept
    value of the best linear fits for different stellar populations in the inner Galaxy with low [Mg/Fe] ratio and R$_{Gal} <$ 6 kpc. Also given are the 16\% and 84\% percentiles of the abundance gradients and of the intercept.}
	\label{low_mgfe_gradients_table}
\begin{tabular}{lcccccc|lcccccc} \hline\hline
\multicolumn{14}{c}{low [Mg/Fe] stars}\\
\hline\hline
\multicolumn{14}{c}{bar stars}\\
\hline
[X/H]&m$_{50}$&m$_{16}$&m$_{84}$&b$_{50}$&b$_{16}$&b$_{84}$&[X/Fe]&m$_{50}$&m$_{16}$&m$_{84}$&b$_{50}$&b$_{16}$&b$_{84}$\\
\hline 
Mg&0.005&$-$0.005&0.015&0.298&0.284&0.312&Mg&$-$0.004&$-$0.006&$-$0.002&0.048&0.045&0.051\\
Si&0.009&$-$0.001&0.020&0.243&0.228&0.258&Si&0.001&$-$0.002&0.003&$-$0.008&$-$0.011&$-$0.005\\
Ca&0.007&$-$0.003&0.017&0.175&0.160&0.190&Ca&$-$0.002&$-$0.004&0.001&$-$0.076&$-$0.080&$-$0.072\\
Al&0.005&$-$0.007&0.017&0.260&0.243&0.277&Al&$-$0.003&$-$0.006&$-$0.000&$-$0.003&$-$0.008&0.001\\
K&0.008&$-$0.005&0.021&0.353&0.335&0.372&K&0.003&$-$0.003&0.008&0.091&0.083&0.098\\
Mn&0.033&0.013&0.054&0.425&0.395&0.454&Mn&0.023&0.015&0.031&0.157&0.145&0.169\\
Co&0.014&$-$0.002&0.030&0.404&0.381&0.427&Co&0.008&0.001&0.015&0.140&0.130&0.150\\
Ni&0.017&0.005&0.029&0.306&0.289&0.323&Ni&0.007&0.004&0.009&0.049&0.046&0.053\\
Fe&0.009&$-$0.002&0.020&0.250&0.234&0.265&Fe&---&---&---&---&---&---\\
Ce*&0.026&$-$0.006&0.057&$-$0.170&$-$0.218&$-$0.121&Ce*&0.032&0.000&0.063&$-$0.387&$-$0.435&$-$0.338\\
Nd*&$-$0.015&$-$0.042&0.012&0.020&$-$0.019&0.061&Nd*&$-$0.017&$-$0.050&0.016&$-$0.187&$-$0.236&$-$0.138\\
\hline
\multicolumn{14}{c}{low ecc stars}\\
\hline
[X/H]&m$_{50}$&m$_{16}$&m$_{84}$&b$_{50}$&b$_{16}$&b$_{84}$&[X/Fe]&m$_{50}$&m$_{16}$&m$_{84}$&b$_{50}$&b$_{16}$&b$_{84}$\\
\hline
Mg&$-$0.012&$-$0.018&$-$0.005&0.244&0.217&0.271&Mg&0.002&0.001&0.004&0.044&0.038&0.050\\
Si&$-$0.012&$-$0.018&$-$0.005&0.206&0.177&0.235&Si&0.002&0.001&0.003&0.004&$-$0.001&0.010\\
Ca&$-$0.007&$-$0.014&$-$0.001&0.131&0.102&0.159&Ca&0.008&0.007&0.010&$-$0.079&$-$0.086&$-$0.073\\
Al&$-$0.015&$-$0.021&$-$0.008&0.220&0.191&0.250&Al&$-$0.000&$-$0.002&0.002&0.008&$-$0.001&0.017\\
K&$-$0.026&$-$0.034&$-$0.017&0.305&0.268&0.342&K&$-$0.010&$-$0.015&$-$0.005&0.090&0.068&0.112\\
Mn&$-$0.026&$-$0.036&$-$0.016&0.368&0.324&0.412&Mn&$-$0.012&$-$0.015&$-$0.008&0.158&0.142&0.174\\
Co&$-$0.022&$-$0.031&$-$0.014&0.365&0.327&0.405&Co&$-$0.005&$-$0.008&$-$0.002&0.145&0.131&0.159\\
Ni&$-$0.016&$-$0.023&$-$0.008&0.260&0.227&0.293&Ni&$-$0.000&$-$0.002&0.001&0.050&0.044&0.056\\
Fe&$-$0.014&$-$0.021&$-$0.007&0.200&0.170&0.230&Fe&---&---&---&---&---&---\\
Ce*&0.039&0.025&0.052&$-$0.223&$-$0.285&$-$0.162&Ce*&0.030&0.019&0.042&$-$0.285&$-$0.338&$-$0.236\\
Nd*&0.010&$-$0.004&0.025&$-$0.103&$-$0.166&$-$0.038&Nd*&0.009&$-$0.006&0.024&$-$0.188&$-$0.256&$-$0.120\\
\hline
\multicolumn{14}{c}{high ecc stars}\\
\hline
[X/H]&m$_{50}$&m$_{16}$&m$_{84}$&b$_{50}$&b$_{16}$&b$_{84}$&[X/Fe]&m$_{50}$&m$_{16}$&m$_{84}$&b$_{50}$&b$_{16}$&b$_{84}$\\
\hline
Mg&$-$0.005&$-$0.012&0.002&0.283&0.269&0.297&Mg&$-$0.005&$-$0.007&$-$0.004&0.051&0.048&0.054\\
Si&0.002&$-$0.006&0.009&0.226&0.212&0.240&Si&0.001&0.000&0.003&$-$0.007&$-$0.009&$-$0.004\\
Ca&0.000&$-$0.007&0.008&0.158&0.144&0.172&Ca&0.000&$-$0.001&0.002&$-$0.077&$-$0.080&$-$0.074\\
Al&$-$0.002&$-$0.012&0.009&0.226&0.205&0.246&Al&$-$0.003&$-$0.005&$-$0.002&$-$0.008&$-$0.012&$-$0.004\\
K&0.014&0.005&0.024&0.312&0.293&0.330&K&0.004&0.001&0.007&0.091&0.085&0.098\\
Mn&0.008&$-$0.007&0.023&0.410&0.380&0.442&Mn&0.005&$-$0.000&0.011&0.173&0.162&0.184\\
Co&0.016&0.005&0.027&0.363&0.342&0.384&Co&0.007&0.003&0.010&0.139&0.132&0.145\\
Ni&0.008&$-$0.001&0.016&0.292&0.275&0.309&Ni&0.004&0.002&0.005&0.054&0.052&0.057\\
Fe&0.001&$-$0.007&0.009&0.231&0.215&0.246&Fe&---&---&---&---&---&---\\
Ce*&0.041&0.023&0.059&$-$0.220&$-$0.260&$-$0.180&Ce*&0.041&0.024&0.059&$-$0.386&$-$0.424&$-$0.348\\
Nd*&0.039&0.023&0.054&$-$0.040&$-$0.071&$-$0.008&Nd*&0.017&$-$0.000&0.033&$-$0.219&$-$0.254&$-$0.184\\
\hline
	\end{tabular}
\end{table*}

\begin{table*}[h]
	\centering
	\scriptsize
	\caption{Radial abundance gradients (dex kpc$^{-1}$) and intercept values of the best linear fits for different stellar populations in the inner Galaxy with high [Mg/Fe] ratio and R$_{Gal} <$ 6 kpc. We also show the 16\% and 84\% percentiles of the abundance gradients and of the intercept.}
	\label{high_mgfe_gradients_table}
\begin{tabular}{lcccccc|lcccccc} \hline\hline
\multicolumn{14}{c}{high [Mg/Fe] stars}\\
\hline\hline
\multicolumn{14}{c}{bar stars}\\
\hline
[X/H]&m$_{50}$&m$_{16}$&m$_{84}$&b$_{50}$&b$_{16}$&b$_{84}$&[X/Fe]&m$_{50}$&m$_{16}$&m$_{84}$&b$_{50}$&b$_{16}$&b$_{84}$\\
\hline 
Mg&0.061&0.047&0.077&$-$0.208&$-$0.228&$-$0.189&Mg&$-$0.015&$-$0.019&$-$0.010&0.300&0.295&0.306\\
Si&0.064&0.049&0.080&$-$0.318&$-$0.337&$-$0.299&Si&$-$0.012&$-$0.017&$-$0.007&0.190&0.184&0.197\\
Ca&0.064&0.049&0.078&$-$0.400&$-$0.419&$-$0.381&Ca&$-$0.012&$-$0.017&$-$0.007&0.107&0.101&0.114\\
Al&0.089&0.069&0.110&$-$0.370&$-$0.397&$-$0.344&Al&0.008&0.002&0.014&0.147&0.139&0.154\\
K&0.070&0.051&0.089&$-$0.250&$-$0.275&$-$0.226&K&$-$0.008&$-$0.015&$-$0.002&0.263&0.255&0.271\\
Mn&0.106&0.080&0.133&$-$0.614&$-$0.650&$-$0.578&Mn&0.020&0.013&0.027&$-$0.073&$-$0.083&$-$0.064\\
Co&0.075&0.055&0.095&$-$0.360&$-$0.386&$-$0.335&Co&$-$0.002&$-$0.007&0.004&0.154&0.147&0.161\\
Ni&0.078&0.061&0.095&$-$0.447&$-$0.469&$-$0.424&Ni&0.001&$-$0.002&0.003&0.064&0.060&0.067\\
Fe&0.076&0.057&0.094&$-$0.510&$-$0.533&$-$0.486&Fe&---&---&---&---&---&---\\
Ce*&0.040&0.010&0.070&$-$0.560&$-$0.597&$-$0.524&Ce*&$-$0.009&$-$0.032&0.014&$-$0.013&$-$0.041&0.016\\
Nd*&0.020&$-$0.013&0.053&$-$0.332&$-$0.373&$-$0.292&Nd*&$-$0.043&$-$0.075&$-$0.012&0.161&0.123&0.199\\
\hline
\multicolumn{14}{c}{low ecc stars}\\
\hline
[X/H]&m$_{50}$&m$_{16}$&m$_{84}$&b$_{50}$&b$_{16}$&b$_{84}$&[X/Fe]&m$_{50}$&m$_{16}$&m$_{84}$&b$_{50}$&b$_{16}$&b$_{84}$\\
\hline
Mg&0.024&0.018&0.030&$-$0.136&$-$0.160&$-$0.1116&Mg&$-$0.004&$-$0.007&$-$0.002&0.280&0.270&0.289\\
Si&0.025&0.019&0.031&$-$0.242&$-$0.267&$-$0.217&Si&$-$0.003&$-$0.005&$-$0.000&0.172&0.163&0.182\\
Ca&0.027&0.020&0.033&$-$0.322&$-$0.348&$-$0.296&Ca&$-$0.001&$-$0.003&0.001&0.090&0.081&0.099\\
Al&0.020&0.013&0.027&$-$0.187&$-$0.218&$-$0.158&Al&$-$0.005&$-$0.008&$-$0.003&0.214&0.202&0.226\\
K&0.019&0.012&0.027&$-$0.131&$-$0.161&$-$0.100&K&$-$0.007&$-$0.011&$-$0.003&0.273&0.257&0.289\\
Mn&0.023&0.013&0.033&$-$0.434&$-$0.476&$-$0.392&Mn&$-$0.001&$-$0.003&0.002&$-$0.036&$-$0.047&$-$0.026\\
Co&0.022&0.014&0.030&$-$0.237&$-$0.270&$-$0.203&Co&$-$0.003&$-$0.006&$-$0.000&0.171&0.160&0.182\\
Ni&0.030&0.022&0.037&$-$0.354&$-$0.383&$-$0.324&Ni&0.002&0.001&0.003&0.057&0.052&0.062\\
Fe&0.029&0.021&0.036&$-$0.417&$-$0.449&$-$0.386&Fe&---&---&---&---&---&---\\
Ce*&0.032&0.020&0.044&$-$0.550&$-$0.602&$-$0.498&Ce*&$-$0.001&$-$0.008&0.006&$-$0.078&$-$0.108&$-$0.049\\
Nd*&0.029&0.017&0.041&$-$0.364&$-$0.415&$-$0.313&Nd*&$-$0.004&$-$0.017&0.010&0.024&$-$0.032&0.079\\
\hline
\multicolumn{14}{c}{high ecc stars}\\
\hline
[X/H]&m$_{50}$&m$_{16}$&m$_{84}$&b$_{50}$&b$_{16}$&b$_{84}$&[X/Fe]&m$_{50}$&m$_{16}$&m$_{84}$&b$_{50}$&b$_{16}$&b$_{84}$\\
\hline
Mg&$-$0.005&$-$0.013&0.002&$-$0.154&$-$0.166&$-$0.142&Mg&$-$0.000&$-$0.002&0.002&0.294&0.291&0.298\\
Si&$-$0.004&$-$0.011&0.003&$-$0.264&$-$0.275&$-$0.252&Si&0.001&$-$0.001&0.004&0.184&0.180&0.188\\
Ca&$-$0.004&$-$0.011&0.003&$-$0.344&$-$0.355&$-$0.332&Ca&0.002&$-$0.001&0.004&0.102&0.098&0.106\\
Al&$-$0.010&$-$0.020&0.000&$-$0.272&$-$0.289&$-$0.256&Al&0.002&$-$0.002&0.005&0.165&0.160&0.171\\
K&$-$0.004&$-$0.013&0.005&$-$0.192&$-$0.208&$-$0.178&K&0.004&0.000&0.007&0.248&0.242&0.254\\
Mn&$-$0.011&$-$0.025&0.002&$-$0.504&$-$0.526&$-$0.481&Mn&$-$0.001&$-$0.004&0.003&$-$0.059&$-$0.065&$-$0.053\\
Co&$-$0.007&$-$0.017&0.002&$-$0.290&$-$0.306&$-$0.274&Co&0.001&$-$0.002&0.004&0.157&0.152&0.161\\
Ni&$-$0.002&$-$0.011&0.006&$-$0.384&$-$0.398&$-$0.370&Ni&0.003&0.001&0.004&0.065&0.063&0.067\\
Fe&$-$0.005&$-$0.014&0.004&$-$0.449&$-$0.463&$-$0.434&Fe&---&---&---&---&---&---\\
Ce*&0.010&$-$0.001&0.022&$-$0.563&$-$0.584&$-$0.543&Ce*&$-$0.019&$-$0.028&$-$0.009&0.022&0.006&0.037\\
Nd*&0.018&0.006&0.031&$-$0.352&$-$0.372&$-$0.332&Nd*&$-$0.037&$-$0.052&$-$0.023&0.179&0.155&0.203\\
\hline
	\end{tabular}
\end{table*}

\section{[X/Fe] radial gradients of the studied bulge-bar populations}

\begin{figure*}[h]
\centering
	\includegraphics[width=13cm]{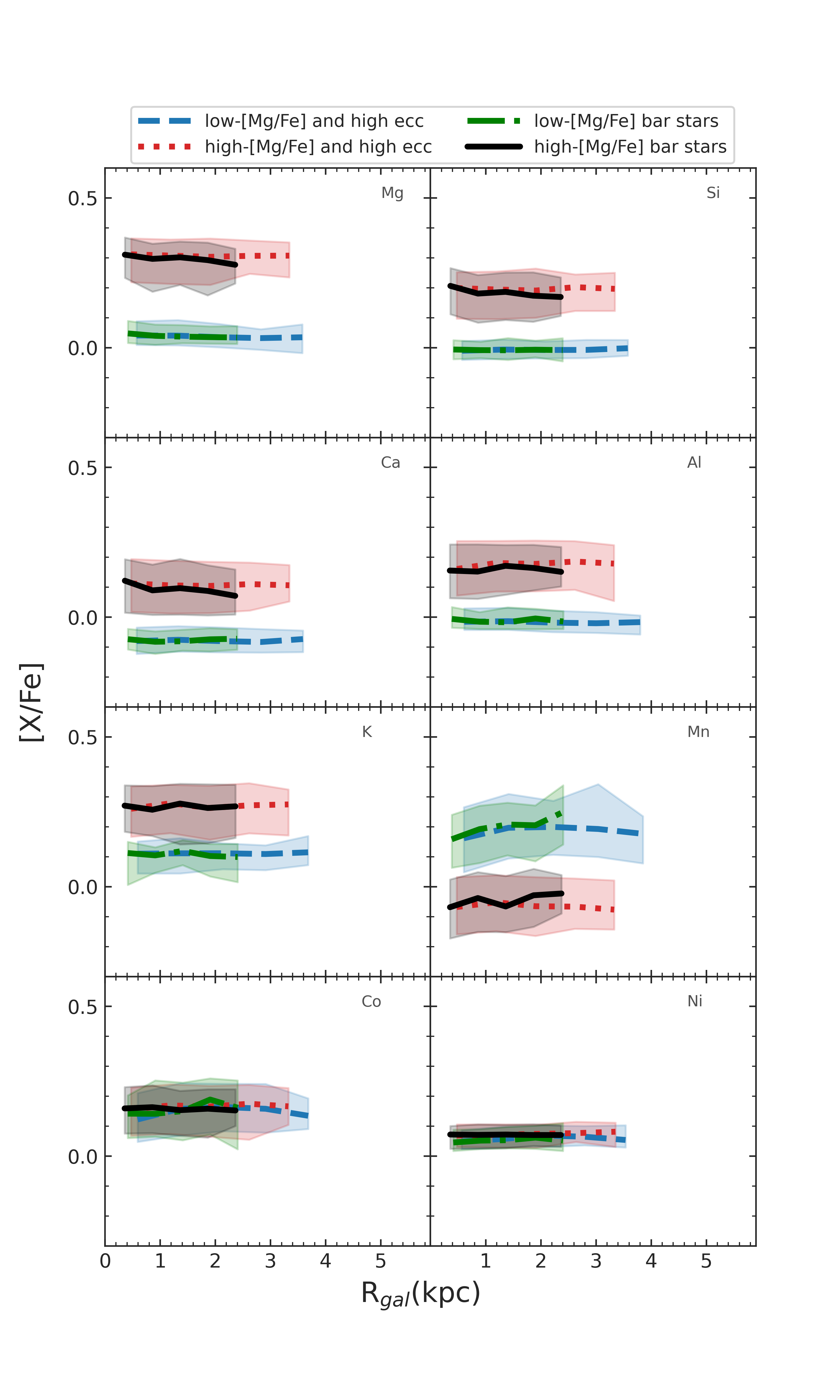}
    \caption{The [X/Fe] abundance ratio as a function of Galactocentric distance for the high-ecc and bar populations. The distinct lines represent the median trends of the bulge-bar populations, with the shaded areas indicating the standard deviation. The legend above the panels describes the meaning of the different lines. 
    }
    \label{X_Fe_vs_rgc_high_ecc}
\end{figure*}

\begin{figure*}[h]
\centering
	\includegraphics[width=13cm]{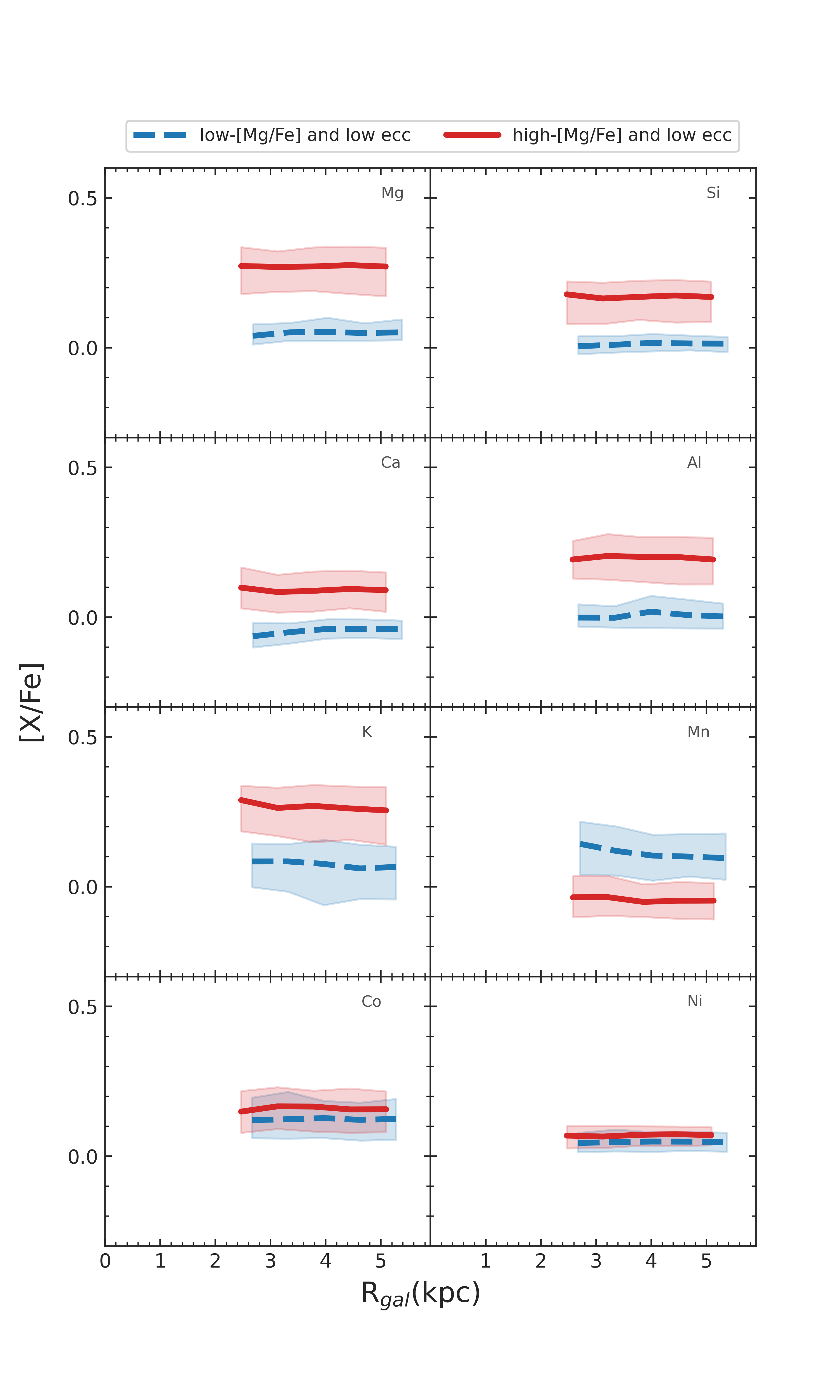}
    \caption{The [X/Fe] abundance ratio as a function of Galactocentric distance for low ecc bulge-bar populations. The distinct lines represent the median trends of the bulge-bar populations, with the shaded areas indicating the standard deviation. The legend above the panels indicates the meaning of the different lines. 
    }
    \label{X_Fe_vs_rgc_low_ecc}
\end{figure*}

\begin{figure*}[h]
\centering
	\includegraphics[width=13cm]{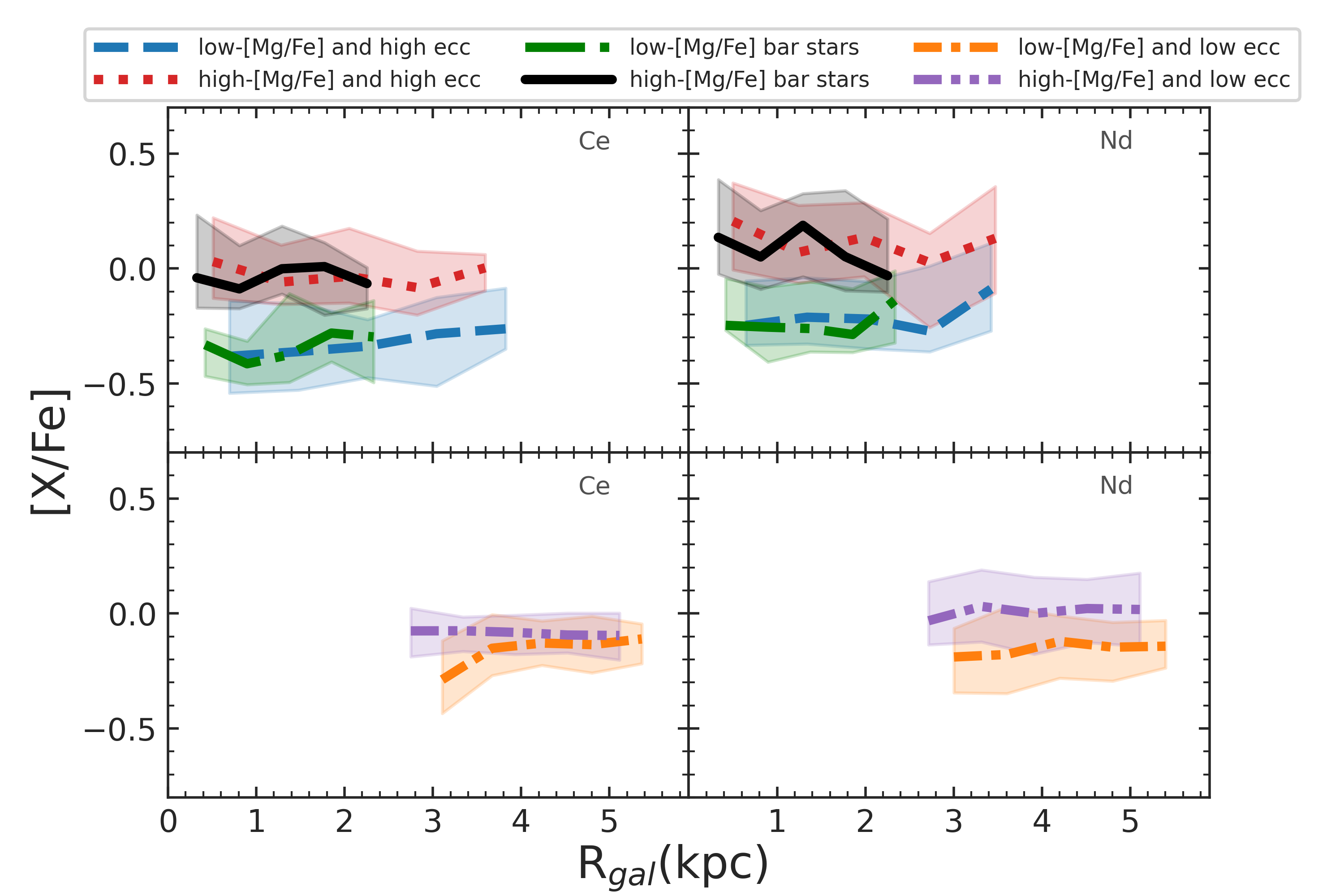}
    \caption{The [Ce/Fe] and [Nd/Fe] abundance ratios as a function of Galactocentric distance for bulge-bar populations using abundances from BAWLAS catalog. The distinct lines represent the median trends, with the shaded areas indicating the standard deviation. The legend above the panels describes the meaning of the different lines. 
    }
    \label{Ce_Nd_Fe_vs_rgc}
\end{figure*}




\bibliography{example}
\bibliographystyle{aasjournalv7}




\end{document}